\documentclass{article}
\usepackage{amsfonts}
\usepackage{amsmath}

\input{tcilatex}
\begin{document}

\setlength{\baselineskip}{20pt}

\begin{center}
\textbf{Double-robust and efficient methods for estimating the causal
effects of a binary treatment.}

\textbf{James Robins}$^{1,3}$\textbf{, Mariela Sued}$^{2}$\textbf{, Quanhong
Lei-Gomez}$^{3}$\textbf{\ and Andrea Rotnitzky}$^{3,4}$
\end{center}

\setlength{\baselineskip}{14pt}

\begin{center}
$^{1}${\small Department of Epidemiology, Harvard School of Public Health}

$^{2}${\small Facultad de Ciencias Exactas y Naturales, Universidad de
Buenos Aires and CONICET,\ Argentina.}

$^{3}${\small Department of Biostatistics, Harvard School of Public Health}

$^{4}\,\,${\small Department of Economics, Di Tella University, Buenos
Aires, Argentina }

\textbf{Abstract}
\end{center}

We consider the problem of estimating the effects of a binary treatment on a continuous outcome of
interest from observational data in the absence of confounding by unmeasured
factors. We provide a new estimator of the population average treatment
effect (ATE) based on the difference of novel double-robust (DR) estimators
of the treatment-specific outcome means. We compare our new estimator with
previously estimators both theoretically and via simulation. DR-difference
estimators may have poor finite sample behavior when the estimated
propensity scores in the treated and untreated do not overlap. We therefore
propose an alternative approach, which can be used even in this unfavorable
setting, based on locally efficient double-robust estimation of a
semiparametric regression model for the modification on an additive scale of
the magnitude of the treatment effect by the baseline covariates $X$. In
contrast with existing methods, our approach simultaneously provides
estimates of: i) the average treatment effect in the total study population,
ii) the average treatment effect in the random subset of the population with
overlapping estimated propensity scores, and iii) the treatment effect at
each level of the baseline covariates $X$.

When the covariate vector $X$ is high dimensional, one cannot be certain,
owing to lack of power, that given models for the propensity score and for
the regression of the outcome on treatment and $X$ used in constructing our
DR estimators are nearly correct, even if they pass standard goodness of fit
tests. Therefore to select among candidate models, we propose a novel
approach to model selection that leverages the DR-nature of our treatment
effect estimator and that outperforms cross-validation in a small simulation
study\footnote{This manuscript was submitted to, and rejected by, the Journal Statistical Science in 2007 and was never revised or submitted elsewhere. However, we are placing it on the arXiv now because of its relevance to a recent important paper by Cui and Tchetgen Tchetgen (2020).}.


\setlength{\parindent}{0.3in} \setlength{\baselineskip}{24pt}

\def\baselinestretch{1.5}\small\normalsize%

\section{Introduction}

In this paper we investigate several approaches to estimation of the causal effects of a binary treatment on a continuous outcome of interest from observational data in the absence of confounding by unmeasured factors.

We first consider estimation of the population ATE. In Section 3 we describe
an approach that treats the problem of estimating the ATE as two missing
data problems, each one regarding the outcomes of subjects that do not
follow a given treatment as missing. We provide novel double-robust (DR)
estimators that are guaranteed to be more efficient than the difference of
inverse probability weighted (IPW) estimators when a model for the
propensity score is correct.

DR-difference estimators will generally have poor small sample behavior
when, as often occurs in practice, there is incomplete overlap of the
estimated propensity scores in the treated and untreated. In this setting,
the present state-of -the-art is to use methods that restrict the analysis
to the random subset of treated and untreated units whose estimated
propensity scores overlap (or can be matched), totally ignoring the data on
the response $Y$ for all units who are not members of the analysis subset.
By restricting the analysis to this random subset of the study population,
these methods may increase both the power (for certain departures from the
null) and validity of tests of the null hypothesis of no treatment effect at
all levels of the baseline covariates $X;$ the drawback is that the causal
contrast being estimated is generally of little or no substantive interest.
Furthermore, these methods suffer from a second important limitation. When a
treatment is of benefit to subjects with certain covariate values but
harmful to others, neither the subset of covariate values for which
treatment is indicated nor the subset for which it is contraindicated will
be discovered using current methods, as these methods average over
covariate-specific treatment effects.

In Section 4 we propose an alternative unified approach that may be more
informative and robust than current methods. Our unified approach is based
on locally efficient double robust estimation of a semiparametric regression
model for the modification on an additive scale of the magnitude of
treatment effect by the baseline covariates $X.$ Our approach, like current
approaches, allows one to test the null hypothesis of no treatment effect at
any level of the baseline covariates $X.$ However, in contrast with current
approaches, our approach simultaneously estimates the average treatment
effect in the total study population, the average treatment effect in the
random subset of the population with overlapping propensity scores, and the
treatment effect at each level of the baseline covariates $X$. We compare
the bias, efficiency, sensitivity to model misspecification and small sample
performance of our unified approach with that of several alternatives
approaches by way of a small simulation study and asymptotic theory.

When the covariate vector $X$ is high dimensional, one cannot be certain,
owing to lack of power, that given models for the propensity score and the
regression of the outcome on treatment and covariates used in constructing
our DR estimators are nearly correct, even if they pass standard goodness of
fit tests. Therefore a number of models for the propensity score with
different subsets of the covariates, different orders of interactions, and
different dimensions of the parameter vector should be fit to the data.
Similarly a number different outcome models should be fit. This raises the
issue that once fit, it is then necessary to select, from among the various
candidate models, the model for the propensity score and the model for the
outcome regression that are to be used in the construction of a DR\
estimator of the average treatment effect. A standard approach would be to
select among candidate models by cross-validation. In section 6, we propose
a novel alternative method of model selection that leverages the DR-nature
of our treatment effect estimator. In a small simulation study, we show that
our proposed method performs well; in particular it outperforms
cross-validation. Our novel method extends to the estimation of any
parameter that admits a DR estimator.

\section{The set-up}

Consider an observational exposure study of the effect of a binary treatment
on a response of interest. For each study subject $i$ we observe the vector $%
O_{i}=\left( T_{i},Y_{i},X_{i}\right) ,$ where $T_{i}$ is a binary treatment
indicator, $Y_{i}$ is the response of interest, higher values of which we
assume are beneficial, and $X_{i}$ is a covariate vector taking values in a
set $\mathcal{X}$. To define causal effects, we conceptualize on each study
subject $i,$ two potential or counterfactual outcomes, namely $Y_{\left(
0\right) i}$ and $Y_{\left( 1\right) i}.$ These are subject's \thinspace$i$
responses to treatment $t=0$ and $t=1$ respectively$.$ We regard the study
subjects as a random sample of size $n$ from a much larger target population
of interest of size $N$. Because $n/N$ \ is small, we may regard the data $%
\left\{ \left( O_{i},Y_{\left( 0\right) ,i},Y_{\left( 1\right) ,i}\right)
;i=1,...,n\right\} $ as $n$ i.i.d. copies of a random vector $\left(
O,Y_{\left( 0\right) },Y_{\left( 1\right) }\right) $ with distribution equal
to the marginal distribution of a single $\left( O_{j},Y_{\left( 0\right)
,j},Y_{\left( 1\right) ,j}\right) $ drawn at random from the target
population. See Robins (1989) for further discussion.

The parameter $\tau_{pop}$ $\equiv\mu_{1}-\mu_{0}$ with $\mu_{t}\equiv
E\left( Y_{(t)}\right) $ $,$ $t=0,1$ is the average treatment effect in the
target population and is therefore of public health interest. The function $%
\gamma(x)\equiv E\left( Y_{\left( 1\right) }|X=x\right) -E\left( Y_{\left(
0\right) }|X=x\right) $ that quantifies the treatment effect among target
population subjects with $X=x\ $is also a parameter of importance when there
is any possibility that treatment may be harmful to some subjects and
beneficial to others, because treatment should be given, since beneficial,
to a subject $i$ with $\gamma\left( X_{i}\right) >0$ but withheld, since
harmful, from a subject $j$ with $\gamma\left( X_{j}\right) <0.$ We note
that ${\tau}_{pop}$ ${=E}\left\{ \gamma(X)\right\} .$ In addition to the
function $\gamma(\cdot)$ and the contrast ${\tau}_{pop},$ the conditional
average treatment effect ${\tau}_{con}\equiv\mathbb{P}_{n}\left\{
\gamma(X)\right\} \equiv n^{-1}\sum_{i=1}^{n}\left\{ \gamma(X_{i})\right\} $
has recently received increased attention in the econometric literature. The
parameter ${\tau}_{con}$ measures the average treatment effect in the study
sample. {The purpose of this paper is to compare various estimators of these
causal contrasts in terms of robustness, efficiency, sensitivity to model
misspecification, and finite sample performance. }We are particularly
interested in settings in which the variance of $\left[ \pi\left( X\right)
\left\{ 1-\pi\left( X\right) \right\} \right] ^{-1}$ is large, where $%
\pi\left( X\right) =P\left( T=1|X\ \right) $ is the propensity score.

Of course, none of the three causal contrasts, $\gamma(x),$ ${\tau}_{pop}$
and ${\tau}_{con}$ is identified from a random sample of the observed data $%
O=\left( T,Y,X\right) $ without further assumptions. To identify these
contrasts we make the following three assumptions that we henceforth assume
to be true.

\textbf{Consistency: }The response to the treatment actually received is the
same as the potential response to that treatment. That is, $Y=Y_{\left(
T\right) }$.

\textbf{Ignorability: }$\left( Y_{(0)},Y_{(1)}\right) $ and $T$ are
conditionally independent given $X$.

\textbf{Strong} \textbf{positivity:} The propensity score $\pi\left(
\cdot\right) \equiv P\left( T=1|X=\cdot\right) $ is bounded away from 0 and
1. That is, $\pi\left( \cdot\right) $ belongs to the set \newline
$\mathcal{P=}\left\{ p\left( \cdot\right) :\text{ }p\left( x\right) \in\left[
\varepsilon,1-\varepsilon\right] \ \text{for all }x\in \mathcal{X}\text{ }%
\right\} $ for some small $\varepsilon>0$ .

We denote by $\mathcal{F}\times\mathcal{P}$ the model characterized by the
above three restrictions plus the restriction that $var\left( Y|X,T\right) $
is bounded with probability 1. Under this model, $E\left( Y_{\left( t\right)
}|X=x\right) =m_{t}\left( x\right) $ where $m_{t}\left( x\right) \equiv
E\left( Y|T=t,X=x\right) $ and thus $\gamma(x)=m_{1}\left( x\right)
-m_{0}\left( x\right) $ and ${\tau}_{pop}=E\left\{ m_{1}\left( X\right)
-m_{0}\left( X\right) \right\} $. $\ $Furthermore, for $t=0,1,$ $%
\mu_{t}=E\left\{ I\left( T=t\right) Y/\pi_{t}\left( X\right) \right\}
=E\left\{ m_{t}\left( X\right) \right\} $ where $\pi_{t}\left( x\right)
\equiv\pi\left( x\right) ^{t}\left( 1-\pi\left( x\right) \right)
^{1-t}=P\left( T=t|X=x\right) $\textbf{\ }

In Section 3, we consider estimating $\tau_{pop}$ as the difference $%
\widehat{\mu}_{1}-\widehat{\mu}_{0}$ of double-robust estimators $\widehat{%
\mu}_{t}$ of the $\mu_{t},$ $t=0,1.$ We construct the $\widehat{\mu }_{t}$
by regarding the estimation of $\mu_{t}$ as a missing data problem in which
the response $Y$ is missing for subjects with $T=1-t.$ We can then construct
double-robust (DR)\ estimators of the $\mu_{t},$ $t=0,1$. These methods use
estimates $\widehat{\pi}\left( x\right) \ $and $\widehat{m}_{t}\left(
x\right) $ of $\pi\left( x\right) $ and $m_{t}\left( x\right) $ based on
parametric models. We discuss methods to make these DR difference estimators
as efficient as possible.

RSGR's results imply that even the most efficient versions of our DR
difference estimators perform poorly when $\left[ \widehat{\pi}\left(
X\right) \left\{ 1-\widehat{\pi}\left( X\right) \right\} \right] ^{-1}$ has
a very large variance. Furthermore, when $\left[ \widehat{\pi}\left(
X\right) \left\{ 1-\widehat{\pi}\left( X\right) \right\} \right] ^{-1}$ is
highly variable, methods based on propensity score matching or
stratification will also fail because there\emph{\ }will usually be
incomplete overlap of the estimated propensity scores for the treated and
untreated. That is, there will be: a) untreated subjects, i.e. with $T=0,$
for whom the value of $\widehat{\pi}\left( X\right) $ will be smaller than
the minimum of $\widehat{\pi}\left( X\right) $ among treated subjects
and/or, b) treated subjects, i.e. with $T=1,$ for whom the value of $%
\widehat{\pi}\left( X\right) $ will be larger than the maximum of $\widehat{%
\pi}\left( X\right) $ among untreated subjects. [In general $\left[ \widehat{%
\pi}\left( X\right) \left\{ 1-\widehat{\pi}\left( X\right) \right\} \right]
^{-1}$ will be highly variable whenever $\left[ \pi\left( X\right) \left\{
1-\pi\left( X\right) \right\} \right] ^{-1}$ itself is highly variable.]
These considerations imply that\emph{\ }when $\left[ \widehat{\pi}\left(
X\right) \left\{ 1-\widehat{\pi}\left( X\right) \right\} \right] ^{-1}$ is highly variable, alternative approaches are needed. A number of authors (LaLonde, 1986, Ho, Imai, King and Stuart, 2005, Heckman, Ichimura and Todd, 1997, Heckman, Ichimura, Smith and Todd, 1998, Smith and Todd, 2007, and Crump, Hotz, Imbens, and Mitnik, 2006) have argued that in this setting, rather than attempting to estimate the average treatment effect parameters $\tau_{pop}$ and $\tau_{con},$ one should estimate either a random contrast such as $\tau_{overlap}=E\left\{ Y_{\left( 1\right)
}-Y_{\left( 0\right) }|\widehat{\pi}\left( X\right) \in\widehat{\mathcal{C}}%
\right\} $ where $\widehat{\mathcal{C}}$ is a random subinterval of $\left[
0,1\right] $ on which the estimated propensity scores for the treated and
untreated overlap, or an analogous contrast based on propensity score
matching. For such contrasts one can often find consistent asymptotically
normal (CAN) estimators that are well behaved in finite samples (i.e. with
small bias and variance) because they disregard the data of units with
estimated propensity scores outside the region of overlap $\widehat{\mathcal{%
C}}$.

The contrasts $\tau_{overlap}$ and $\tau_{pop}$ are but two examples of a $%
\omega-$weighted population average treatment effect parameter $\tau
_{pop,\omega}=E\left\{ \omega\left( X\right) \gamma\left( X\right) \right\}
/E\left\{ \omega\left( X\right) \right\} $ where $\omega\left( \cdot\right) $
is non-negative, the first having $\omega\left( x\right) =I\left( \widehat{%
\pi}\left( x\right) \in\widehat{\mathcal{C}}\right) $ and the second having $%
\omega\left( x\right) =1$ for all $x$. Similarly, $\tau_{con}$ is an example
of the $\omega-$weighted conditional average treatment effect parameter $%
\tau_{con,\omega}=\mathbb{P}_{n}\left\{ \omega\left( X\right)
\gamma(X)\right\} /\mathbb{P}_{n}\left\{ \omega\left( X\right) \right\} $ in
which $\omega\left( x\right) =1$ for all $x.$ However, except for a few
important exceptions, parameters such as $\tau_{overlap}$ in which $%
\omega\left( \cdot\right) $ is a function of the empirical distribution of
the estimated propensity score have little substantive meaning; quantifying
the average treatment effect in a random subpopulation is often of limited
scientific or public health value. [An example of such an exception is an
observational study of an already licensed drug in which the goal is to
estimate the effect on the subpopulation of patients for whom consensus as
to the value of the drug does not yet exist, as evidenced by estimated
propensity scores far from either zero or one.] In the non-exceptional
cases, the reason for restricting attention to causal contrasts on a random
subpopulation is to increase both the power and validity of tests of the $%
\gamma-$null hypothesis 
\begin{equation}
\gamma\left( X\right) =0\text{ w.p.1}  \label{null}
\end{equation}
of no treatment effect at any level of $X.$ To see this note that the
contrasts $\tau_{pop,\omega}$ and $\tau_{con,\omega}$ are 0 under $\left( %
\ref{null}\right) .$ Thus, an $\alpha$ level test of $\tau_{pop,\omega}=0$ \
(or of $\tau_{con,\omega}=0),$ is also an $\alpha$ level test of the null
hypothesis $\left( \ref{null}\right) $. For certain random weights such as $%
\omega\left( x\right) =I\left( \widehat{\pi}\left( x\right) \in \widehat{%
\mathcal{C}}\right) ,$ one can often construct tests of $\tau_{con,\omega}=0$
or of $\tau_{pop,\omega}=0$ with level close to nominal and with reasonable
power for detecting departures from the null $\left( \ref{null}\right) $ for 
$X^{\prime}s\,\ $in $\widehat{\pi}^{-1}\left( \widehat{\mathcal{C}}\right) $
(although with no power to detect departures for $X^{\prime}s\,\ $not in $%
\widehat{\pi}^{-1}\left( \widehat{\mathcal{C}}\right) ).$

Indeed Crump et. al. (2006) argued that if an analyst is willing to change
the parameter of interest from ${\tau}_{pop}$ to some $\tau_{pop,\omega}$ or 
$\tau_{con,\omega}$ so as to avoid the difficulties in estimating ${\tau }%
_{pop}$ and $\tau_{con}$, he should perhaps choose $\omega\left(
\cdot\right) $ to be the one$\ $for which $\tau_{con,\omega}$ is easiest to
estimate. They then proved that, under homoscedascity, i.e. when $var\left(
Y|X,T\right) =\sigma^{2},$ the semiparametric variance bound for estimation
of $\tau_{con,\omega}$ in our model $\mathcal{F}\times\mathcal{P}$ is $\left[
{E}\left\{ \omega\left( X\right) \right\} \right] ^{-2}E\left[ \omega\left(
X\right) ^{2}/\left[ \pi\left( X\right) \left\{ 1-\pi\left( X\right)
\right\} \right] \right] $ which is minimized at $\left[ E\left\{
\omega_{op}\left( X\right) \right\} \right] ^{-1}$ over all weight functions 
$\omega\left( X\right) $ where 
\begin{equation}
\omega_{op}\left( X\right) =\pi\left( X\right) \left\{ 1-\pi\left( X\right)
\right\} .  \label{wop}
\end{equation}

In Section 4 we present an approach that differs from that proposed by the
aforementioned authors. Our approach, like theirs, can often provide an $%
\alpha$-level test of the $\gamma-$null hypothesis with good power when $%
\left[ \widehat{\pi}\left( X\right) \left\{ 1-\widehat{\pi}\left( X\right)
\right\} \right] ^{-1}$\emph{\ }is highly variable. However, our approach,
in contrast with the others, additionally delivers\textbf{\ }estimates of $%
\tau_{pop},$ $\tau_{con}$ and the treatment function $\gamma\left(
\cdot\right) ,$ as well as of the contrasts $\tau_{con,\omega }$ and $%
\tau_{pop,\omega}$ estimated by the other approaches. Our unified approach
is based on fitting a semiparametric regression model for the modification,
on an additive scale, of the magnitude of treatment effect by the baseline
covariates $X$. Because such semiparametric regression model is indeed a
structural nested mean model specialized to a point exposure study, our
approach adheres to the general recommendation given by Robins (1999) that
it is usually advantageous to use structural nested models rather than
marginal structural models to estimate treatment effects.

\section{Double-robust difference estimators}

Our double-robust difference estimators are constructed as follows. Given a
sequence $v_{l}\left( X\right) ,l=1,2,...$ of specified functions of $X$
which are dense in $L_{2}\left( F_{X}\right) $ ($F_{X}$ being the marginal
law of $X)\,\ $and such that $v_{1}\left( X\right) =1,$ we postulate the
working outcome models%
\begin{equation}
m_{t}\left( x\right) \equiv E\left( Y|T=t,X=x\right) =\Phi^{-1}\left\{
t\varsigma_{1}^{\ast T}v^{k}\left( x\right) +\left( 1-t\right)
\varsigma_{0}^{\ast T}v^{k}\left( x\right) \right\}  \label{total}
\end{equation}
where $\Phi\left( \cdot\right) $ is some suitable canonical link function, $%
v^{d}\left( x\right) \equiv\left( v_{1}\left( x\right) ,...,v_{d}\left(
x\right) \right) ^{T}$ for any $d,$ and $\varsigma_{t}^{\ast}$ is an unknown
conformable parameter vector.

We also postulate the working parametric propensity score model 
\begin{equation}
\pi\left( x\right) =\pi\left( x;\alpha^{\ast}\right)  \label{prop-mod}
\end{equation}
where $\pi\left( \cdot;\cdot\right) $ is a known function and $\alpha^{\ast
} $ is an unknown Euclidean vector, (for example, $\pi\left( x;\alpha\right)
=$ \newline
expit$\left\{ \alpha^{T}v^{s}\left( x\right) \right\} $ for some $s)$ and we
estimate $\alpha^{\ast}$ with its maximum likelihood estimator $\widehat{%
\alpha}.$ We define $\widehat{\pi}\left( x\right) \equiv\pi\left( x;\widehat{%
\alpha}\right) $ and $\widehat{\pi}_{t}\left( x\right) \equiv\widehat{\pi}%
\left( x\right) ^{t}\left\{ 1-\widehat{\pi }\left( x\right) \right\} ^{1-t}$.

Next, with $\widehat{m}_{t}\left( x\right) =\Phi^{-1}\left\{ \widehat {%
\varsigma}_{t}^{T}v^{k}\left( x\right) \right\} $ where $\widehat {\varsigma}%
_{t}^{T}$ is some CAN estimator of $%
\varsigma_{t}^{\ast}$ under model $\left( \ref{total}\right) ,$ we compute $%
\widehat{\mu}_{t,B-DR}\left( \widehat{m}_{t},\widehat{\pi }\right) $ where,
for any $m_{t}\left( \cdot\right) $ and $\pi\left( \cdot\right) ,$ 
\begin{equation}
\widehat{\mu}_{t,B-DR}\left( m_{t},\pi\right) \equiv\mathbb{P}_{n}\left\{
m_{t}\left( X\right) \right\} +\frac{\mathbb{P}_{n}\left[ \frac{I\left(
T=t\right) }{\pi_{t}\left( X\right) }\left\{ Y-m_{t}\left( X\right) \right\} %
\right] }{\mathbb{P}_{n}\left\{ \frac{I\left( T=t\right) }{\pi_{t}\left(
X\right) }\right\} },t=0,1  \label{dr-est}
\end{equation}
and, for any $W,$ $\mathbb{P}_{n}\left( W\right) =\sum_{i=1}^{n}W_{i}/n.$
Finally, we estimate $\tau_{pop}$ and $\tau_{con}$ with the difference $%
\widehat{\tau}_{B-DR}\left( \widehat{m},\widehat{\pi}\right) =\widehat{\mu }%
_{1,B-DR}\left( \widehat{m}_{1},\widehat{\pi}\right) -\widehat{\mu }%
_{0,B-DR}\left( \widehat{m}_{0},\widehat{\pi}\right) $.

The estimator $\widehat{\tau}_{B-DR}\left( \widehat{m},\widehat{\pi}\right) $
is double-robust for $\tau_{pop}$ and $\tau_{con}$ in the union submodel of $%
\mathcal{F}\times\mathcal{P}$, throughout denoted as $\mathcal{U}_{0}\left( 
\mathcal{F}\times\mathcal{P}\right) ,$ which makes the assumption that model 
$\mathcal{F}\times\mathcal{P}$ and either (but not necessarily both) the
propensity score model $\left( \ref{prop-mod}\right) $ or the outcome model $%
\left( \ref{total}\right) $ is correct, i.e. $\sqrt{n}\left\{ \widehat{\tau}%
_{B-DR}\left( \widehat{m},\widehat{\pi}\right) -\tau _{pop}\right\} $ and $%
\sqrt{n}\left\{ \widehat{\tau}_{B-DR}\left( \widehat{m},\widehat{\pi}\right)
-\tau_{con}\right\} $ converge to a mean zero normal distribution when model 
$\mathcal{F}\times\mathcal{P}$ holds and at least one of $\left( \ref%
{prop-mod}\right) $ or $\left( \ref{total}\right) $ is correct.

A number of options are available for CAN estimation of $m_{t}\left(
x\right) .$ An obvious first choice is to estimate it with $\widehat {m}%
_{t,REG}\left( x\right) =\Phi^{-1}\left\{ \widehat{\varsigma}%
_{t,REG}^{T}v^{k}\left( x\right) \right\} $ where $\widehat{\varsigma }%
_{t,REG}^{T}$ is the iteratively reweighted least squares (IWLS) estimator
of $\varsigma_{t}^{\ast T}$ satisfying 
\begin{equation*}
\mathbb{P}_{n}\left[ I\left( T=t\right) \left\{ Y-\Phi^{-1}\left( \widehat{%
\varsigma}_{t,REG}^{T}V^{\dagger}\right) \right\} V^{\dagger }\right] =0, 
\end{equation*}
and where, throughout, $V^{\dagger}\equiv v^{k}\left( X\right) .$ When $Y$
is discrete the estimator $\widehat{\tau}_{B-DR}\left( \widehat{m}_{REG},%
\widehat{\pi}\right) $ has the drawback that it is not guaranteed to lie in
the parameter space for $\tau_{pop}.$ An alternative estimation strategy for
the regression functions $m_{t}\left( x\right) ,t=0,1,$ remedies this
problem. Specifically, consider estimating $m_{t}\left( x\right) $ with the
weighted least squares estimator $\widehat{m}_{t,WLS}\left( x\right)
=\Phi^{-1}\left( \widehat{\varsigma}_{t,WLS}^{T}v^{k}\left( x\right) \right) 
$ where $\widehat{\varsigma}_{t,WLS}^{T}$ satisfies\emph{\ }%
\begin{equation}
\mathbb{P}_{n}\left[ \frac{I\left( T=t\right) }{\widehat{\pi}_{t}\left(
X\right) }\left\{ Y-\Phi^{-1}\left( \widehat{\varsigma}_{t,WLS}^{T}V^{%
\dagger}\right) \right\} V^{\dagger}\right] =0,\text{ \ \ \ }t=0,1.
\label{sep-est}
\end{equation}
As pointed out by Robins, Sued, Lei-Gomez and Rotnitzky (2007) (throughout
referred to as RSGR), $\left( \ref{sep-est}\right) $ implies that $\widehat{%
\tau}_{B-DR}\left( \widehat{m}_{WLS},\widehat{\pi}\right) $ is equal to $%
\mathbb{P}_{n}\left\{ \widehat{m}_{t,WLS}\left( X\right) \right\} .$ Thus,
with an adequately chosen link function, $\widehat{\tau }_{B-DR}\left( 
\widehat{m}_{WLS},\widehat{\pi}\right) $ is guaranteed to lie in the
parameter space for $\tau_{pop}.$ We also consider a DR-difference estimator 
$\widehat{\tau}_{B-DR}\left( \widehat{m}_{NR},\widehat{\pi}\right) $ that is
also guaranteed to lie in the parameter space for $\tau_{pop}$ for an
appropriately chosen link function. To compute $\widehat{m}_{t,NR}\left(
x\right) $ we consider the extended model $m_{t}\left( x\right) =\Phi
^{-1}\left\{ \varsigma_{t}^{\ast T}v^{k}\left( x\right) +\phi_{t}^{\ast }%
\widehat{\pi}_{t}\left( x\right) \right\} $ and we estimate $\varsigma _{t}$
and $\phi_{t}$ with the weighted least squares estimators $\widehat {%
\varsigma}_{t,NR}$ and $\widehat{\phi}_{t,NR}$ satisfying%
\begin{equation*}
\mathbb{P}_{n}\left[ \frac{I\left( T=t\right) }{\widehat{\pi}_{t}\left(
X\right) }\left\{ Y-\Phi^{-1}\left( \widehat{\varsigma}_{t,NR}^{T}V^{%
\dagger}+\widehat{\phi}_{t,NR}^{T}\widehat{\pi}_{t}\left( X\right) \right)
\right\} \left( 
\begin{array}{c}
V^{\dagger} \\ 
\widehat{\pi}_{t}\left( X\right)%
\end{array}
\right) \right] =0,\text{ \ \ \ }t=0,1 
\end{equation*}
Because we have assumed the vector $V^{\dagger}$ has one component equal to
the constant $1$, $\widehat{\tau}_{B-DR}\left( \widehat{m}_{NR},\widehat{\pi 
}\right) =\mathbb{P}_{n}\left( \widehat{m}_{1,NR}\right) -\mathbb{P}%
_{n}\left( \widehat{m}_{0,NR}\right) $. Furthermore, since by construction, $%
\mathbb{P}_{n}\left[ I\left( T=t\right) \left\{ Y-\widehat{m}_{t,NR}\left(
X\right) \right\} \right] =0,$ we find that $\widehat{\mu }_{t,DR}\left( 
\widehat{m}_{t,NR},\widehat{\pi}\right) $ is also equal to $\mathbb{P}%
_{n}\left\{ I\left( T=t\right) Y+I\left( T=1-t\right) \widehat{m}%
_{t,NR}\left( X\right) \right\} .$Thus the estimator $\widehat{\tau}%
_{B-DR}\left( \widehat{m}_{NR},\widehat{\pi}\right) \ $has the property that
the outcome model $\left( \emph{\ref{total}}\right) $ is only used to impute
each subject's unobserved counterfactual outcome $Y_{1-T}.$ As a consequence 
$\widehat{\tau}_{B-DR}\left( \widehat{m}_{NR},\widehat{\pi}\right) $ is
likely to be less sensitive than the estimator $\widehat{\tau}_{B-DR}\left( 
\widehat{m}_{WLS},\widehat{\pi }\right) $ to misspecification of model $%
\left( \ref{total}\right) $ when model $\left( \ref{prop-mod}\right) $ is
incorrect.

In the Appendix we show that when both $\left( \ref{prop-mod}\right) $ and $%
\left( \ref{total}\right) $ are correct, 
\begin{equation*}
var^{A}\left\{ \sqrt{n}\left( \widehat{\tau}_{B-DR}\left( \widehat {m},%
\widehat{\pi}\right) -\tau_{con}\right) \right\} =\sigma^{2}E\left\{
\omega_{op}\left( X\right) ^{-1}\right\} 
\end{equation*}
and 
\begin{equation}
var^{A}\left\{ \sqrt{n}\left( \widehat{\tau}_{B-DR}\left( \widehat {m},%
\widehat{\pi}\right) -\tau_{pop}\right) \right\} =\sigma^{2}E\left\{
\omega_{op}\left( X\right) ^{-1}\right\} +var\left\{ \gamma\left( X\right)
\right\}  \label{vardr}
\end{equation}
where here and throughout, for any sequence $W_{n},var^{A}\left(
W_{n}\right) $ stands for the variance of the limiting distribution of $%
W_{n}.$ Indeed, it follows from Rotnitzky and Robins (1995), Robins and
Ritov (1997), and Crump, et.al. (2006), that these asymptotic
variances are the semiparametric variance bounds for estimators of $%
\tau_{con}$ and $\tau_{pop}$ respectively in model $\mathcal{F}\times%
\mathcal{P}$ (and in its union submodel $\mathcal{U}_{0}\left( \mathcal{F}%
\times\mathcal{P}\right) $) when both $\left( \ref{prop-mod}\right) $ and $%
\left( \ref{total}\right) $ hold. That is the estimators $\widehat{\tau}%
_{B-DR}\left( \widehat {m},\widehat{\pi}\right) $ are locally semiparametric
efficient in the model $\mathcal{F}\times\mathcal{P}$ (and in its union
submodel $\mathcal{U}_{0}\left( \mathcal{F}\times\mathcal{P}\right) $) at
the intersection submodel in which both $\left( \ref{prop-mod}\right) $ and $%
\left( \ref{total}\right) $ happen to be true.

\subsection{Double-robust estimators of the treatment specific means
guaranteed to be at least as efficient than IPW estimators when the
propensity model is correct.}

When a model for the propensity score $\pi\left( x\right) $ is fit by
maximum likelihood, $\widehat{\mu}_{t,B-DR}\left( \widehat{m}_{t},\widehat{%
\pi}\right) $ is not only double-robust but also never less (and generally
more) efficient in large samples than the inverse weighted probability
estimator $\widehat{\mu}_{t,IPW}$ and the Horvitz-Thompson like estimator $%
\widehat{\mu}_{t,HT},$ 
\begin{equation*}
\widehat{\mu}_{t,IPW}\equiv\left. \mathbb{P}_{n}\left\{ \frac{I\left(
T=t\right) }{\widehat{\pi}_{t}\left( X\right) }Y\right\} \right/ \mathbb{P}%
_{n}\left\{ \frac{I\left( T=t\right) }{\widehat{\pi}_{t}\left( X\right) }%
\right\} \text{ and }\widehat{\mu}_{t,HT}\equiv\mathbb{P}_{n}\left\{ \frac{%
I\left( T=t\right) }{\widehat{\pi}_{t}\left( X\right) }Y\right\} ,t=0,1 
\end{equation*}
whenever the propensity model $\left( \ref{prop-mod}\right) $ and the
outcome model $\left( \ref{total}\right) $ are both correct. However, as
shown in the Appendix, the estimator $\widehat{\mu}_{t,B-DR}\left( \widehat{m%
}_{t},\widehat{\pi}\right) $ has the drawback that it may be less efficient
than either $\widehat{\mu}_{t,IPW}$ or $\widehat{\mu}_{t,HT}$ if the
propensity score model is correct but the outcome model is misspecified.

To overcome this drawback, we now construct two DR estimators of $\mu_{t}$, 
\newline
$\widehat{\mu}_{t,B-DR}\left( \widehat{m}_{t,REG},\widehat{\pi }%
_{ITER-REG}^{\left( t\right) }\right) $ and $\widehat{\mu}_{t,B-DR}\left( 
\widehat{m}_{t,ITER-WLS},\widehat{\pi}_{ITER-WLS}\right) ,$ which, when the
propensity score model is correct, are guaranteed to be i) at least as
efficient as $\widehat{\mu}_{t,B-DR}\left( \widehat{m}_{t,REG},\widehat{\pi }%
\right) $ and $\widehat{\mu}_{t,B-DR}\left( \widehat{m}_{t,WLS},\widehat{\pi}%
\right) $ respectively and, ii) at least as efficient as the most efficient
of\emph{\ }$\widehat{\mu}_{t,IPW}$\emph{\ }and\emph{\ }$\widehat{\mu}_{t,HT}$
. In addition, like $\widehat{\mu}_{t,B-DR}\left( \widehat{m}_{t,WLS},%
\widehat{\pi}\right) ,$ the estimator $\widehat{\mu}_{DR}\left( \widehat{m}%
_{t,ITER-WLS},\widehat{\pi}_{ITER-WLS}^{\left( t\right) }\right) $ is a
regression estimator, i.e. it is equal to $\mathbb{P}_{n}\left\{ \widehat{m}%
_{t,ITER-WLS}\left( X\right) \right\} $ and hence guaranteed to lie in the
parameter space of $\mu_{t}.$ Robins, Rotnitzky and Bonetti (2001), Robins
(2002) and van der Laan and Rubin (2006) previously described iterative
procedures which, as the number of iterations goes to infinity, converge to
a DR estimator of $\mu_{t}$ that satisfy property ii) and are guaranteed to
fall in the parameter space for $\mu_{t}$. The estimator $\widehat{\mu}%
_{DR}\left( \widehat{m}_{t,ITER-WLS},\widehat {\pi}_{ITER-WLS}^{\left(
t\right) }\right) $ satisfies these same properties after a single
iteration. Tan (2006) proposed a non-recursive DR estimator satisfying
property ii) but which, in contrast to $\widehat{\mu }_{DR}\left( \widehat{m}%
_{t,ITER-WLS},\widehat{\pi}_{ITER-WLS}^{\left( t\right) }\right) ,$ is not
guaranteed to fall in the parameter space for $\mu_{t}$ when this space is
not the entire real line.

For concreteness, we shall describe our DR estimators assuming $\widehat{\mu 
}_{t,IPW}$ and $\widehat{\mu}_{t,HT}$ use $\widehat{\pi}\left( x\right) $
computed from the ML fit of model $\left( \ref{prop-mod}\right) $ where 
\begin{equation}
\pi\left( x;\alpha\right) =\text{expit}\left\{ \alpha^{T}q\left( x\right)
\right\}  \label{st-prop}
\end{equation}
and $q\left( x\right) $ is a conformable vector of specified, i.e. known,
functions of $x.$ Extension to general propensity score models $\left( \ref%
{prop-mod}\right) $ is carried out by replacing in what follows each
instance of $q\left( x\right) $ with $\partial$logit$\left\{ \pi\left(
x;\alpha\right) \right\} /\partial\alpha$.

The estimator $\widehat{\mu}_{t,B-DR}\left( \widehat{m}_{t,ITER-WLS},%
\widehat{\pi}_{ITER-WLS}^{\left( t\right) }\right) $ uses the fitted
regression and propensity functions $\widehat{m}_{t,ITER-WLS}\left( x\right) 
$ and $\widehat{\pi}_{ITER-WLS}^{\left( t\right) }\left( x\right) ,$ $t=0,1,$
computed as follows. We first compute $\widehat{m}_{t,WLS}\left( x\right)
,t=0,1,$ and the fitted values $\widehat{\pi}\left( x\right) $ from the ML\
fit of model $\left( \ref{st-prop}\right) $ as indicated above. Having done
so, we compute the (data-dependent) functions $\widehat{g}_{t,1}\left(
x\right) =\widehat{m}_{t,WLS}\left( x\right) /\widehat{\pi }_{t}\left(
x\right) $ and $\widehat{g}_{t,2}\left( x\right) =1/\widehat{\pi}_{t}\left(
x\right) ,t=0,1$. Next, we formulate extended logistic regression models
that specify that $\pi\left( x\right) =\pi_{ITER-WLS}^{\left( t\right)
}\left( x;\alpha_{t}^{\ast},\varphi
_{t,1}^{\ast},\varphi_{t,2}^{\ast}\right) $ where 
\begin{equation}
\pi_{ITER-WLS}^{\left( t\right) }\left(
x;\alpha_{t},\varphi_{t,1},\varphi_{t,2}\right) =\text{expit}\left\{
\alpha_{t}^{T}q\left( x\right) +\varphi_{t,1}\widehat{g}_{t,1}\left(
x\right) +\varphi_{t,2}\widehat {g}_{t,2}\left( x\right) \right\} ,t=0,1,
\label{wls-iter}
\end{equation}
and we compute the ML estimators $\left( \widehat{\alpha}_{t,ITER-WLS},%
\widehat{\varphi}_{t,1,ITER-WLS},\widehat{\varphi}_{t,2,ITER-WLS}\right) $
of $\left( \alpha_{t},\varphi_{t,1},\varphi_{t,2}\right) ,$\newline
$t=0,1,$ under the corresponding model. We then define $\widehat{\pi}%
_{ITER-WLS}^{\left( t\right) }\left( x\right) =$\newline
$\pi_{ITER-WLS}^{\left( t\right) }\left( x;\widehat{\alpha}_{t,ITER-WLS},%
\widehat{\varphi }_{t,1,ITER-WLS},\widehat{\varphi}_{t,2,ITER-WLS}\right) $.
Finally we compute \newline
$\widehat{m}_{t,ITER-WLS}\left( x\right) =\Phi^{-1}\left( \widehat{\varsigma}%
_{t,ITER-WLS}^{T}v^{k}\left( x\right) \right) $ where $\widehat{\varsigma}%
_{t,ITER-WLS}^{T}$ satisfies 
\begin{equation*}
\mathbb{P}_{n}\left[ \frac{I\left( T=t\right) }{\widehat{\pi}%
_{t,ITER-WLS}^{\left( t\right) }\left( X\right) }\left\{ Y-\Phi ^{-1}\left( 
\widehat{\varsigma}_{t,ITER-WLS}^{T}V^{\dagger}\right) \right\} V^{\dagger}%
\right] =0,\text{ \ \ \ }t=0,1 
\end{equation*}

The estimator $\widehat{\mu}_{t,B-DR}\left( \widehat{m}_{t,REG},\widehat{\pi 
}_{ITER-REG}^{\left( t\right) }\right) $ uses $\widehat{m}_{t,REG}\left(
x\right) $ defined as above and $\widehat{\pi}_{ITER-REG}^{\left( t\right)
}\left( x\right) $ computed just like $\widehat{\pi}_{ITER-WLS}^{\left(
t\right) }\left( x\right) $ except with $\widehat{g}_{t,1}\left( x\right) $
redefined as $\widehat{m}_{t,REG}\left( x\right) /\widehat{\pi}_{t}\left(
x\right) .$

In the Appendix we sketch the proof that the estimators $\widehat{\mu }%
_{t,B-DR}\left( \widehat{m}_{t,REG},\widehat{\pi}_{ITER-REG}^{\left(
t\right) }\right) $ and $\widehat{\mu}_{t,B-DR}\left( \widehat {m}%
_{t,ITER-WLS},\widehat{\pi}_{ITER-WLS}^{\left( t\right) }\right) ,t=0,1,$
satisfy the aforementioned desirable properties i) and ii). However, the
following warning is appropriate at this point: there is no guarantee that
the desirable property i) is retained when the propensity score model is
incorrect and the outcome model is correct. Indeed, in additional simulation
experiments not reported here, we observed that the relative efficiencies
reversed under this scenario. Thus, for now, pending further theoretical and
Monte Carlo investigations, we would only recommend routine use of the
iterative propensity double-robust estimators developed in this section in
settings in which the selected propensity model is more likely to be nearly
correct than the selected outcome model. An example might be a setting in
which (i)\ the propensity and outcome model were chosen from among many
candidates by a procedure (e.g. cross validation) that minimizes estimated
mean squared prediction error for $T$ and $Y$ respectively and (ii) the
outcome $Y$ was a Bernoulli with a small success probability, so the
effective sample size for selecting the outcome model is small.

\section{An approach based on modeling the modification of treatment effect
by baseline covariates.}

\ We will now consider the unified approach mentioned in the introduction
that delivers simultaneous estimates of $\tau_{pop},$ $\tau_{con}$ and the
treatment function $\gamma\left( \cdot\right) ,$ as well as of the contrasts 
$\tau_{con,\omega}$ and $\tau_{pop,\omega}$. In this section, we assume that
the response $Y$ is continuous. Our unified approach produces a CAN
estimator $\widehat{\beta}^{T}v^{d}\left( x\right) $ of an approximation to
the function $\gamma\left( x\right) ,$ namely its projection (with regards
to a specific inner product defined later) into the space generated by the
first $d$ elements of the sequence $v_{l}\left( x\right) ,l=1,2,...,$ for
some user-specified integer $d$. The estimator $\widehat{\beta}^{T}$ is
defined in Equation $\left( \ref{betahat}\right) $ below.

When the semiparametric regression (SR) model%
\begin{equation}
\gamma\left( x\right) =\beta^{\ast T}v^{d}\left( x\right) \text{ }
\label{gamma}
\end{equation}
holds, $\gamma\left( X\right) $ equals its projection and consequently, $%
\widehat{\beta}^{T}v^{d}\left( X\right) $ is a CAN estimator of $%
\gamma\left( X\right) $ and $\mathbb{P}_{n}\left\{ \omega\left( X\right) 
\widehat{\beta}^{T}V\right\} /\mathbb{P}_{n}\left\{ \omega\left( X\right)
\right\} \ $is a CAN estimator of $\tau_{pop,\omega}$ and $\tau_{con,\omega}$
for arbitrary $\omega.$ We have called model $\left( \ref{gamma}\right) $ a
semiparametric regression model because it is equivalent to the model
\begin{equation}
E\left( Y|T=t,X=x\right) =t\beta^{\ast T}v^{d}\left( x\right) +m_{0}\left(
x\right)  \label{SRM}
\end{equation}
in which $m_{0}\left( \cdot\right) =E\left( Y|T=0,X=\cdot\right) $ is
unspecified.

When the SR model $\left( \ref{gamma}\right) $ does not hold, our estimator $%
\widehat{\beta}^{T}v^{d}\left( x\right) $ converges in probability to $\beta^{\dagger T}v^{d}\left( x\right) $ with 
\begin{equation}
\beta^{\dagger}\equiv E_{\omega_{op}}\left\{ VV^{T}\right\}
^{-1}E_{\omega_{op}}\left\{ \gamma\left( X\right) V\right\} ,
\label{betatarget}
\end{equation}
where, throughout,\emph{\ }$V\equiv v^{d}\left( X\right) $\emph{\ }and $%
E_{\omega}\left( W\right) $ stands for $E\left\{ \omega\left( X\right)
W\right\} /E\left\{ \omega\left( X\right) \right\} .$ The quantity\emph{\ }$%
\beta^{\dagger T}$\emph{\ }is the minimizer of\emph{\ }$\left\Vert
\gamma\left( X\right) -\beta^{T}V\right\Vert _{\omega}^{2}\equiv E_{\omega }%
\left[ \left\{ \gamma\left( X\right) -\beta^{T}V\right\} ^{2}\right] $\emph{%
\ }with $\omega=\omega_{op}$ over all vectors\emph{\ }$\beta,$ i.e. $%
\beta^{\dagger T}V$ is equal to the projection (denoted throughout as $%
\Pi_{\omega}\left( \gamma\left( X\right) |\left[ V\right] \right) $) of $%
\gamma\left( X\right) $ into the first $d$ components of the sequence $%
v_{l}\left( X\right) ,l=1,2,...,$ with respect to the inner product $%
\left\langle W_{1},W_{2}\right\rangle _{\omega}\equiv E_{\omega}\left(
W_{1}W_{2}\right) $ with $\omega=\omega_{op}.$ We shall see below that the
specific form $\omega_{op}$ for the weight $\omega$ arises because $\widehat{%
\beta}$ is chosen so that $\widehat{\beta}^{T}v^{d}\left( x\right) $ is
locally semiparametric efficient for $\gamma\left( x\right) $ under the SR
model $\left( \ref{gamma}\right) $ at a specific submodel. Note that when
model $\left( \ref{gamma}\right) $ holds, $\beta^{\dagger}$ is equal to $%
\beta^{\ast}$ and $\beta^{\dagger T}v^{d}\left( x\right) $ is equal to $%
\gamma\left( x\right) .$

More specifically, the estimator $\widehat{\beta}$ defined by $\left( \ref%
{betahat}\right) $ below is double-robust for the parameter $\beta^{\dagger}$
defined in $\left( \ref{betatarget}\right) $ in a union submodel $\mathcal{U}%
_{1}\left( \mathcal{F}\times\mathcal{P}\right) $ of $\mathcal{F}\times%
\mathcal{P}$ different from $\mathcal{U}_{0}\left( \mathcal{F}\times\mathcal{%
P}\right) .$ Like $\mathcal{U}_{0}\left( \mathcal{F}\times\mathcal{P}\right)
,$ model $\mathcal{U}_{1}\left( \mathcal{F}\times\mathcal{P}\right) $ makes
the assumption that one of the components of the union is the propensity
score model $\left( \ref{prop-mod}\right) .$ However, $\mathcal{U}_{1}\left( 
\mathcal{F}\times\mathcal{P}\right) $ differs from $\mathcal{U}_{0}\left( 
\mathcal{F}\times \mathcal{P}\right) $ in that the second component of the
union is a model, throughout referred to as the $H-$ model, that postulates
that 
\begin{equation}
E\left\{ H\left( \beta^{\dagger}\right) |X=x\right\} =\theta^{\ast
T}v^{k}\left( x\right)  \label{reg-mod}
\end{equation}
for some unknown Euclidean vector $\theta^{\ast}$, where for any $\beta,$ $%
H\left( \beta\right) \equiv Y-T\beta^{T}V$ and recall, $V^{\dagger}\equiv
v^{k}\left( X\right) $ for some $k$ which we choose greater than or equal to 
$d,$ making $V$ a subvector of $V^{\dagger}.$

If indeed model $\left( \ref{gamma}\right) $ holds so that $\beta^{\ast
}=\beta^{\dagger}$, the $H-$model $\left( \ref{reg-mod}\right) $ is the same
as the model $m_{0}\left( X\right) =\theta^{\ast T}V^{\dagger}$ because,
under $\left( \ref{gamma}\right) ,$ $E\left\{ H\left( \beta^{\ast}\right)
|X\right\} =m_{0}\left( X\right) =E\left( Y|T=0,X\right) .$ That is, if
model $\left( \ref{gamma}\right) $ holds, the $H-$model $\left( \ref{reg-mod}%
\right) $ is equivalent to the earlier outcome model $\left( \ref{total}%
\right) $ restricted to $t=0$ and $\Phi$ the identity.

Henceforth for ease of reference, we call $\mathcal{F}_{sub}\times\mathcal{P}
$ the submodel of $\mathcal{F}\times\mathcal{P}$ which, in addition to the
assumptions encoded in $\mathcal{F}\times\mathcal{P}$, assumes the SR model$%
\left( \ref{gamma}\right) $ holds.

Because under $\left( \ref{gamma}\right) ,$ $\beta^{\ast}=\beta^{\dagger},$
an estimator $\widehat{\beta}$ which is double-robust for $\beta^{\dagger}$
in model $\mathcal{U}_{1}\left( \mathcal{F}\times\mathcal{P}\right) ,$ is
automatically double-robust for $\beta^{\ast}$ in the union submodel $%
\mathcal{U}_{1}\left( \mathcal{F}_{sub}\times\mathcal{P}\right) $ of $%
\mathcal{F}_{sub}\times\mathcal{P}$ in which either the propensity score
model is correct or $E\left( Y|T=0,X\right) =\theta^{\ast T}V^{\dagger}$ for
some $\theta^{\ast}.$ An estimator $\widehat{\beta}$ double-robust in $%
\mathcal{U}_{1}\left( \mathcal{F}_{sub}\times\mathcal{P}\right) $ renders
the estimators $\widehat{\beta}^{T}v^{d}\left( x\right) $ and $\mathbb{P}%
_{n,\omega}\left( \widehat{\beta}^{T}V\right) $ double-robust for $%
\gamma\left( x\right) $ and $\tau_{pop,\omega}$ (and $\tau_{con,\omega}),$
respectively, in $\mathcal{U}_{1}\left( \mathcal{F}_{sub}\times \mathcal{P}%
\right) ,$ where henceforth, for any function $W=w\left( O\right) $ of the
observed data, $\mathbb{P}_{n,\omega}\left( W\right) \equiv\mathbb{P}%
_{n}\left\{ \omega\left( X\right) W\right\} /\mathbb{P}_{n}\left(
\omega\left( X\right) \right) .$

We now provide the definition of our estimator $\widehat{\beta}$ and argue
that it is double-robust for $\beta^{\dagger}$ in model $\mathcal{U}%
_{1}\left( \mathcal{F}\times\mathcal{P}\right) .$ Specifically $\widehat{%
\beta}\ $is the first component of the ternary $\left( \widehat{\beta},%
\widehat{\theta},\widehat{\alpha}\right) $ satisfying $\mathbb{P}_{n}\left\{
S\left( \widehat{\beta},\widehat{\alpha},\widehat{\theta}\right) \right\} =0$
where 
\begin{equation*}
S\left( \beta,\alpha,\theta\right) \equiv\left( 
\begin{array}{c}
S_{1}\left( \beta,\alpha,\theta\right) \\ 
S_{2}\left( \beta,\theta\right) \\ 
S_{3}\left( \alpha\right)%
\end{array}
\right) \equiv\left( 
\begin{array}{c}
V\left\{ H\left( \beta\right) -\theta^{T}V^{\dagger}\right\} \left\{
T-\pi\left( X;\alpha\right) \right\} \\ 
V^{\dagger}\left\{ H\left( \beta\right) -\theta^{T}V^{\dagger}\right\} \\ 
\left\{ \partial\text{ logit}\pi\left( X;\alpha\right) /\partial
\alpha\right\} \left\{ T-\pi\left( X;\alpha\right) \right\}%
\end{array}
\right) 
\end{equation*}
Some algebra gives 
\begin{equation}
\widehat{\beta}=\mathbb{P}_{n}\left[ V\left\{ T-\pi\left( X;\widehat {\alpha}%
\right) \right\} \Pi_{n}\left[ TV^{T}|\left[ V^{\dagger}\right] ^{\perp}%
\right] \right] ^{-1}\mathbb{P}_{n}\left[ V\left\{ T-\pi\left( X;\widehat{%
\alpha}\right) \right\} \Pi_{n}\left[ Y|\left[ V^{\dagger }\right] ^{\perp}%
\right] \right]  \label{betahat}
\end{equation}
where, with $W_{j}=w\left( O_{j}\right) $ denoting any function of the $%
j^{th}$ sample unit observed data, $j=1,..,n,$ $\Pi_{n}\left( W|\left[
V^{\dagger}\right] ^{\perp}\right) _{j}$ stands for $W_{j}-\mathbb{P}%
_{n}\left( WV^{\dagger T}\right) \mathbb{P}_{n}\left( V^{\dagger}V^{\dagger
T}\right) ^{-1}V_{j}^{\dagger}$ with $V_{j}^{\dagger}=v^{k}\left(
X_{j}\right) .$

Under regularity conditions, $\sqrt{n}\left\{ \left( \widehat{\beta },%
\widehat{\alpha},\widehat{\theta}\right) -\left( \beta^{\dagger\dagger
},\alpha^{\dagger},\theta^{\dagger}\right) \right\} $ converges in law to a
mean zero normal distribution, where $\left(
\beta^{\dagger\dagger},\alpha^{\dagger},\theta^{\dagger}\right) $ solves $%
E\left\{ S\left( \beta,\alpha,\theta\right) \right\} =0.$ Regardless of
whether or not model $\left( \ref{gamma}\right) $ for $\gamma\left( x\right) 
$ holds, we show in the Appendix, that when either $\left( \ref{prop-mod}%
\right) $ is correct or model $\left( \ref{reg-mod}\right) $ is correct, $%
\beta^{\dagger\dagger}$ is equal to the parameter $\beta^{\dagger}$ defined
in $\left( \ref{betatarget}\right) $. Thus, in particular when model $\left( %
\ref{gamma}\right) $ for $\gamma\left( x\right) $ is correct, $%
\beta^{\dagger\dagger}=\beta^{\ast}.$ This then shows that $\widehat{\beta}$
is double-robust for $\beta^{\dagger}$ in the union model $\mathcal{U}%
_{1}\left( \mathcal{F}\times\mathcal{P}\right) $ and double-robust for $%
\beta^{\ast}$ in $\mathcal{U}_{1}\left( \mathcal{F}_{sub}\times \mathcal{P}%
\right) $. Furthermore, under $\mathcal{U}_{1}\left( \mathcal{F}_{sub}\times%
\mathcal{P}\right) $ we also have that i) $\sqrt {n}\left( \widehat{\beta}%
^{T}v^{d}\left( x\right) -\gamma\left( x\right) \right) ,$ ii) $\sqrt{n}%
\left\{ \mathbb{P}_{n}\left( \widehat{\beta}^{T}V\right) -\tau_{pop}\right\} 
$ and iii) $\sqrt{n}\left\{ \mathbb{P}_{n}\left( \widehat{\beta}^{T}V\right)
-\tau_{con}\right\} $, are all asymptotically normal with mean 0.

The preceding estimator $\widehat{\beta}$ is not the only DR estimator of $%
\beta^{\ast}$ in $\mathcal{U}_{1}\left( \mathcal{F}_{sub}\times \mathcal{P}%
\right) $. Indeed, given any non-singular user-specified $d\times d$ matrix
function $c\left( x\right) $ of $x$, if $S_{\left( c\right) }\left(
\beta,\alpha,\theta\right) $ is defined like $S\left( \beta
,\alpha,\theta\right) $ except that its first sub-vector $S_{1,\left(
c\right) }\left( \beta,\alpha,\theta\right) $ is defined as $c\left(
X\right) S_{1}\left( \beta,\alpha,\theta\right) ,$ it is straightforward to
show that the $\widehat{\beta}\left( c\right) $-component of the ternary $%
\left( \widehat{\beta}\left( c\right) ,\widehat{\alpha},\widehat{\theta }%
\left( c\right) \right) $ satisfying $\mathbb{P}_{n}\left\{ S_{\left(
c\right) }\left( \widehat{\beta}\left( c\right) ,\widehat{\alpha},\widehat{%
\theta}\left( c\right) \right) \right\} =0$ is double-robust for $%
\beta^{\ast}$ in $\mathcal{U}_{1}\left( \mathcal{F}_{sub}\times \mathcal{P}%
\right) $. The choice of $c\left( \cdot\right) $ impacts the efficiency with
which we estimate $\beta^{\ast}$. The optimal choice depends on what the
true observed data law is. If for the true observed data law, $var\left(
Y|T,X\right) $ is a constant $\sigma^{2}$, and both the propensity score and
the $H$-models hold, then it can be shown that the optimal choice for $%
c\left( X\right) $ is the constant identity matrix $I$. In fact, with this
choice the asymptotic variance of $\widehat{\beta}$ attains the
semiparametric variance bound for estimators of $\beta^{\ast}$ both in model 
$\mathcal{F}_{sub}\times\mathcal{P}$ and in its union submodel $\mathcal{U}%
_{1}\left( \mathcal{F}_{sub}\times\mathcal{P}\right) $. Until Section 4.4 we
will restrict attention to $\widehat{\beta}$ and the associated estimators $%
\widehat{\beta}^{T}v^{d}\left( x\right) $ and $\mathbb{P}_{n}\left( \widehat{%
\beta}^{T}V\right) $. In Section 4.4 we will consider the estimators $%
\widehat{\beta}\left( c\right) $ when we discuss options for reducing bias
in the estimation of $\tau_{pop}$ and $\tau_{con}$ when, in fact, the
semiparametric model $\left( \ref{gamma}\right) $ does not hold.

\subsection{Comparison of $\mathbb{P}_{n}\left( \widehat{\protect\beta}%
^{T}V\right) $ with $\widehat{\protect\tau}_{B-DR}.$}

We now examine how $\widehat{\tau}_{B-DR}\left( \widehat{m},\widehat{\pi }%
\right) $ and $\mathbb{P}_{n}\left( \widehat{\beta}^{T}V\right) $ compare as
estimators of $\tau_{pop}.$ Since our discussion here holds regardless of
the choice of estimators $\widehat{m}_{t}\left( x\right) $ so long as they
are CAN for $m_{t}\left( x\right) $ under model $\left( \ref{total}\right) ,$
in\emph{\ }what follows, we simplify the notation and we let $\widehat{\tau }%
_{B-DR}$ and $\widehat{\mu}_{t,B-DR}$ stand for $\widehat{\tau}_{B-DR}\left( 
\widehat{m},\widehat{\pi}\right) $ and\emph{\ }$\widehat{\mu}_{t,B-DR}\left( 
\widehat{m}_{t},\widehat{\pi}\right) $ respectively. In the computation of
the estimator $\widehat{\tau}_{B-DR}$ treated subjects, i.e. those with $T=1$%
, who have $\widehat{\pi}\left( X\right) $ close to 0 have enormous leverage
on $\widehat{\mu}_{1,B-DR}$ while untreated subjects, i.e. those with $T=0,$
who have $\widehat{\pi}\left( X\right) $ close to 1 have enormous leverage
on $\widehat{\mu}_{0,B-DR}.$ In contrast, these subjects have bounded
influence on $\widehat{\beta},$ and hence on $\mathbb{P}_{n}\left( \widehat{%
\beta}^{T}V\right) ,$ because $\widehat{\pi}\left( X\right) $ enters in the
computation of $\widehat{\beta}$ in $\left( \ref{betahat}\right) $ via the
factor $\left\{ T-\pi\left( X;\widehat{\alpha}\right) \right\} $ which has
absolute value less than 1. This is one important reason why the variance of 
$\widehat{\tau}_{B-DR}$\ is generally, although not always, much greater
than the variance of $\mathbb{P}_{n}\left( \widehat{\beta}^{T}V\right) .$\
In Section 4.3, we describe an unusual setting in which the two estimators
have the same variance.

The price paid for the variance reduction is that, compared to $\widehat{%
\tau }_{B-DR},$ $\mathbb{P}_{n}\left( \widehat{\beta}^{T}V\right) $ is
inconsistent, for $\tau_{pop}$ under more model misspecification scenarios.
Specifically, when model $\left( \ref{gamma}\right) $ for $\gamma\left(
X\right) $ holds, both $\mathbb{P}_{n}\left( \widehat{\beta}^{T}V\right) $
and $\widehat{\tau}_{B-DR}$ are consistent for $\tau_{pop}$ provided either
(but not necessarily both) the propensity score model $\left( \ref{prop-mod}%
\right) $ or the $H-$model $\left( \ref{reg-mod}\right) $ are correct. The
following argument proves the consistency of $\widehat{\tau}_{B-DR}$ under
model $\left( \ref{gamma}\right) $ and the $H-$model $\left( \ref{reg-mod}%
\right) .$ Specifically, the validity of these two models is equivalent to
the validity of the model%
\begin{equation}
E\left( Y|T=t,X=x\right) =t\beta^{\ast T}v^{d}\left( x\right) +\theta^{\ast
T}v^{k}\left( x\right)  \label{OLS}
\end{equation}
\ Because we have assumed that the dimension $k$ of $v^{k}\left(
x\right) $ is no less than the dimension $d$ of $v^{d}\left( x\right) ,$ the
truth of $\left( \ref{OLS}\right) $ implies the truth of\emph{\ }$\left( \ref%
{total}\right) .$ Thus, $\widehat{\tau}_{B-DR}$ is CAN for $\tau_{pop}$
under $\left( \ref{gamma}\right) $ and $\left( \ref{reg-mod}\right) $
because we have already seen that $\widehat{\tau}_{B-DR}$ is CAN for $%
\tau_{pop}$when model $\left( \ref{total}\right) $ holds$.$\emph{\ }In
contrast, when model $\left( \ref{gamma}\right) $ for $\gamma\left( X\right) 
$ is incorrect, $\mathbb{P}_{n}\left( \widehat{\beta}^{T}V\right) $ is
inconsistent for $\tau_{pop}$ but $\widehat{\tau}_{B-DR}$ still holds a
chance of being consistent; specifically, $\widehat{\tau}_{B-DR}$ remains
consistent if the propensity score model is correct or, when $k\neq d,$ if $%
\left( \ref{total}\right) $ is correct (note that the validity of model $%
\left( \ref{total}\right) $ does not imply the validity of model $\left( \ref%
{OLS}\right) $ except when\emph{\ }$k=d)$\emph{.} In conclusion, $\mathbb{P}%
_{n}\left( \widehat{\beta}^{T}V\right) $ is often much superior to $\widehat{%
\tau}_{B-DR}$ as estimator of $\tau_{pop}$ when model $\left( \ref{gamma}%
\right) $ for $\gamma\left( X\right) $ is correct; however, its performance
compared with that of $\widehat{\tau}_{B-DR}$ may degrade when model $\left( %
\ref{gamma}\right) $ is misspecified. We emphasize that if, as we assume
here, the first component of $V$ is $v_{1}\left( X\right) =1,$ then model $%
\left( \ref{gamma}\right) $ is guaranteed correct whenever $\gamma\left(
x\right) $ is a constant function; in particular, it is correctly specified
under the $\gamma-$null hypothesis $\left( \ref{null}\right) .$

\subsection{Comparison of $\mathbb{P}_{n}\left( \widehat{\protect\beta}%
^{T}V\right) $ with estimators based on OLS\ estimation of $\protect\beta%
^{\ast}.$}

Another natural estimator of $\tau_{pop}$ is $\mathbb{P}_{n}\left( \widehat{%
\beta}_{OLS}^{T}V\right) $ where $\left( \widehat{\beta}_{OLS},\widehat{%
\theta}_{OLS}\right) $ are the OLS estimators of $\left(
\beta^{\ast},\theta^{\ast}\right) $ in model $\left( \ref{OLS}\right) .$ The
estimators $\mathbb{P}_{n}\left( \widehat{\beta}_{OLS}^{T}V\right) $ and $%
\mathbb{P}_{n}\left( \widehat{\beta}^{T}V\right) $ are algebraically
identical when the propensity score model assumes that $\pi\left( x\right) $
is constant . Except for this situation, the estimator $\mathbb{P}_{n}\left( 
\widehat{\beta}^{T}V\right) $ has one important robustness property compared
to $\mathbb{P}_{n}\left( \widehat{\beta}_{OLS}^{T}V\right) .$ Specifically,
under the $\gamma-$null hypothesis, when the $H-$model is incorrect, $%
\mathbb{P}_{n}\left( \widehat{\beta}^{T}V\right) $\emph{\ }is consistent if
the chosen propensity score model is correct. In contrast, under the same
scenario, $P_{n}\left( \widehat{\beta}_{OLS}^{T}V\right) $ is generally
inconsistent for $\tau_{pop}=0$ unless the propensity score is a linear
combination of the components of $V^{\dagger}$ w.p.1 (Robins, Mark
and Newey, 1992). The price paid for this robustness is that when the $H-$%
model is correct, the asymptotic variance of $\mathbb{P}_{n}\left( \widehat{%
\beta }_{OLS}^{T}V\right) $ is less than or equal, and generally strictly
smaller than, the asymptotic variance of\emph{\ }$\mathbb{P}_{n}\left( 
\widehat{\beta }^{T}V\right)$.

\subsection{Double-robustness of $\mathbb{P}_{n}\left( \widehat{\protect\beta%
}^{T}V\right) $ as an estimator of $\protect\tau_{pop}$ and $\protect\tau%
_{con}$ under a specific union submodel $\mathcal{U}_{0}\left( \mathcal{F}%
\times \mathcal{P}\right) $ of $\mathcal{F}\times\mathcal{P}$}

We have seen that $\mathbb{P}_{n}\left( \widehat{\beta}^{T}V\right) $ is
double-robust for $\tau_{pop}$ and $\tau_{con}$ in $\mathcal{U}_{1}\left( 
\mathcal{F}_{sub}\times\mathcal{P}\right) $. Suppose that $d=k$\emph{\ }so%
\emph{\ }$V=V^{\dagger}.$ Then, interestingly, when $\pi\left(
x;\alpha^{\ast}\right) $ in model $\left( \ref{prop-mod}\right) $ is%
\begin{equation*}
\pi\left( x;\alpha^{\ast}\right) =\frac{1}{2}\left\{ 1+\sqrt{1-\left(
\alpha^{\ast T}v^{d}\left( x\right) \right) ^{-1}}\right\} 
\end{equation*}
or equivalently, when $\omega_{op}\left( x;\alpha^{\ast}\right) \equiv
\pi\left( x;\alpha^{\ast}\right) \left\{ 1-\pi\left( x;\alpha^{\ast }\right)
\right\} $ satisfies 
\begin{equation}
\left\{ \omega_{op}\left( x;\alpha^{\ast}\right) \right\} ^{-1}=\alpha^{\ast
T}v^{d}\left( x\right)  \label{wop-mod}
\end{equation}
it just happens that $\mathbb{P}_{n}\left( \widehat{\beta}^{T}V\right) $ is
also double-robust for $\tau_{pop}$ and $\tau_{con}$ in the same union model
as $\widehat{\tau}_{B-DR}$,\ i.e. in model $\mathcal{U}_{0}\left( \mathcal{F}%
\times\mathcal{P}\right) $. That is, $\sqrt{n}\left\{ \mathbb{P}_{n}\left( 
\widehat{\beta}^{T}V\right) -\tau_{pop}\right\} $ and $\sqrt{n}\left\{ 
\mathbb{P}_{n}\left( \widehat{\beta}^{T}V\right) -\tau_{con}\right\} $
converge to a mean zero normal distribution under model $\mathcal{F}\times%
\mathcal{P}$ when one, but not necessarily both, of the models $\left( \ref%
{prop-mod}\right) $ (with the specific form $\left( \ref{wop-mod}\right) )$
or $\left( \ref{total}\right) $ is correct. To see why this is the case,
write $\widehat{\omega}_{op}\left( X\right) =\widehat{\pi}\left( X\right)
\left\{ 1-\widehat{\pi}\left( X\right) \right\} $ and $\widehat{\pi}\left(
X\right) =\pi\left( X;\widehat{\alpha }\right) $, and multiply $0=\mathbb{P}%
_{n}\left\{ S_{1}\left( \widehat{\beta},\widehat{\alpha},\widehat{\theta}%
\right) \right\} $ by $\widehat{\alpha}^{T}$ to conclude that $\mathbb{P}_{n}%
\left[ \left\{ \widehat{\omega}_{op}\left( X\right) \right\} ^{-1}\left\{
H\left( \widehat{\beta}\right) -\widehat{\theta}^{T}V^{\dagger}\right\}
\left\{ T-\widehat{\pi}\left( X\right) \right\} \right] =0.$ Next note that $%
\left\{ \widehat{\omega}_{op}\left( X\right) \right\} ^{-1}\left\{ T-%
\widehat{\pi}\left( X\right) \right\} =T\left\{ \widehat{\pi}\left( X\right)
\right\} ^{-1}-\left( 1-T\right) \left( 1-\widehat{\pi}\left( X\right)
\right) ^{-1}$ and conclude that $\mathbb{P}_{n}\left[ \left( Y-\widehat{%
\beta}^{T}V-\widehat{\theta}^{T}V^{\dagger}\right) \left\{ T/\widehat{\pi}%
\left( X\right) \right\} -\left( Y-\widehat{\theta}^{T}V^{\dagger}\right)
\left\{ \left( 1-T\right) /\left( 1-\widehat{\pi }\left( X\right) \right)
\right\} \right] =0$. Finally, deduce that 
\begin{align*}
\mathbb{P}_{n}\left( \widehat{\beta}^{T}V\right) & =\mathbb{P}_{n}\left\{ 
\widetilde{m}_{1,REG}\left( X\right) \right\} +\mathbb{P}_{n}\left[ \frac{T}{%
\widehat{\pi}\left( X\right) }\left\{ Y-\widetilde{m}_{1,REG}\left( X\right)
\right\} \right] \\
& -\left\{ \mathbb{P}_{n}\left\{ \widetilde{m}_{0,REG}\left( X\right)
\right\} +\mathbb{P}_{n}\left[ \frac{1-T}{1-\widehat{\pi}\left( X\right) }%
\left\{ Y-\widetilde{m}_{0,REG}\left( X\right) \right\} \right] \right\}
\end{align*}
where $\widetilde{m}_{t,REG}\left( X\right) =t\widehat{\beta}V+\widehat {%
\theta}^{T}V^{\dagger},$ $t=0,1.$ We therefore see that $\mathbb{P}%
_{n}\left( \widehat{\beta}^{T}V\right) $ is equal to $\widehat{\tau}%
_{B-DR}\left( \widehat{m},\widehat{\pi}\right) $ with $\widehat{m}_{t}\left(
x\right) =\widetilde{m}_{t,REG}\left( x\right) $. But $\widetilde{m}%
_{t,REG}\left( x\right) $ is CAN for $m_{t}\left( x\right) $ under model $%
\left( \ref{total}\right) $ (and consequently $\mathbb{P}_{n}\left( \widehat{%
\beta}^{T}V\right) $ is double-robust in model $U_{0}\left( \mathcal{F}\times%
\mathcal{P}\right) )$ because $\widetilde {m}_{t,REG}\left( x\right) $ is
CAN for $m_{t}\left( x\right) $ under model $\left( \ref{OLS}\right) $ and
the equality of the dimensions $d$ of $V$ and $k$ of $V^{\dagger}$ implies
that model $\left( \ref{OLS}\right) $ holds when model $\left( \ref{total}%
\right) $ holds.

\subsection{Performance under misspecification of the model for $\protect%
\gamma\left( X\right) .$}

We shall now examine the performance, i.e. the (asymptotic) bias and
variance, of $\widehat{\beta}^{T}v^{d}\left( x\right) $ as an estimator of $%
\gamma(x)\ $and of $\mathbb{P}_{n}\left( \widehat{\beta}^{T}V\right) \ $as
an estimator of ${\tau}_{pop}$ or ${\tau}_{con}$ under model $\mathcal{U}%
_{1}\left( \mathcal{F}\times\mathcal{P}\right) ,$ i.e. when model $\left( %
\ref{gamma}\right) $ for $\gamma\left( x\right) $ may be incorrect but at
least one of models $\left( \ref{prop-mod}\right) $ or $\left( \ref{reg-mod}%
\right) $ is correct.

Examination of the asymptotic bias is straightforward since we have already
seen earlier that $\widehat{\beta}$ converges in probability to $\beta
^{\dagger}$ defined in Eq.$\left( \ref{betatarget}\right) $ under model $%
\mathcal{U}_{1}\left( \mathcal{F}\times\mathcal{P}\right) .$ Thus, under $%
\mathcal{U}_{1}\left( \mathcal{F}\times\mathcal{P}\right) ,$ the probability
limits of: i) $\widehat{\beta}^{T}v^{d}\left( x\right) -\gamma\left(
x\right) $ and of ii) $\mathbb{P}_{n}\left( \widehat{\beta }^{T}V\right) -{%
\tau}_{pop}$ and $\mathbb{P}_{n}\left( \widehat{\beta}^{T}V\right)
-\tau_{con},$ are $\beta^{\dagger T}v^{d}\left( x\right) -\gamma\left(
x\right) $ and $\beta^{\dagger T}E\left( V\right) -{\tau }_{pop}$
respectively. Interestingly, in the Appendix we show that 
\begin{equation}
\beta^{\dagger T}E\left( V\right) -{\tau}_{pop}=-E_{\omega_{op}}\left\{
\Pi_{\ \omega_{op}}\left[ \gamma\left( X\right) |\left[ V\right] ^{\perp }%
\right] \Pi_{\ \omega_{op}}\left[ Q|\left[ V\right] ^{\perp}\right] \right\}
\label{ts}
\end{equation}
where $Q=\left[ E\left\{ \omega_{op}\left( X\right) \right\} /\omega
_{op}\left( X\right) \right] -1$ and for any $W,\Pi_{\ \omega_{op}}\left[ W|%
\left[ V\right] ^{\perp}\right] =W-\Pi_{\ \omega_{op}}\left[ W|\left[ V%
\right] \right] .$ In the Appendix we demonstrate that this expression
implies the double-robustness, discussed in Section 4.3, of $\mathbb{P}%
_{n}\left( \widehat{\beta}^{T}V\right) $ as estimator of $\tau_{con}$ and of 
$\tau_{pop}$ in model $\mathcal{U}_{0}\left( \mathcal{F}\times\mathcal{P}%
\right) $ when model $\left( \ref{wop-mod}\right) $ holds and the dimensions 
$d$ and $k$ are equal.

For an arbitrary $\omega\left( \cdot\right) ,$ the asymptotic bias of $%
\mathbb{P}_{n,\omega}\left( \widehat{\beta}^{T}V\right) $ as estimator of ${%
\tau}_{pop,\omega}$ in model $\mathcal{U}_{1}\left( \mathcal{F}\times%
\mathcal{P}\right) \,\,$is equal to $\beta^{\dagger T}E_{\omega}\left(
V\right) -{\tau}_{pop,\omega}.$ Interestingly, in the Appendix we show that
when $\omega=\omega_{op},$ {the asymptotic bias is 0, i.e. }%
\begin{equation}
\beta^{\dagger T}E_{\omega_{op}}\left( V\right) ={\tau}_{pop,\omega_{op}}
\label{wbias}
\end{equation}
However, $\mathbb{P}_{n,\omega_{op}}\left( \widehat{\beta}^{T}V\right) $ is
not a feasible estimator except when $d=1$ (and thus $V=1)$, because, for $%
d>1,$ calculation of $E_{\omega_{op}}\left( V\right) $ requires input of $%
\omega_{op}\left( X\right) $ which is unknown (although consistently
estimable when the propensity model is correct).

The formulae for the asymptotic variance of our estimators under model $%
\mathcal{U}_{1}\left( \mathcal{F}\times\mathcal{P}\right) $ are messy and
are given in the Appendix. However, this formulae simplify to interesting
and interpretable quantities when both models $\left( \ref{prop-mod}\right) $
and $\left( \ref{reg-mod}\right) $ hold and $var\left( Y|X,T\right)
=\sigma^{2}$. The following theorem provides the variance formulae for this
special case. Let $\Sigma=\left\{ E_{\omega_{op}}\left( VV^{T}\right)
\right\} ^{-1},$ $\Omega=\left[ E\left\{ \omega_{op}\left( X\right) \right\} %
\right] ^{-1}$ and $b\left( X\right) =1-3\pi\left( X\right) +3\pi\left(
X\right) ^{2}.$

\textbf{Theorem 1:} Under model $\mathcal{U}_{1}\left( \mathcal{F}\times%
\mathcal{P}\right) $ and when $\left( \ref{prop-mod}\right) $ and $\left( %
\ref{reg-mod}\right) $ hold,

a) $\sqrt{n}\left( \widehat{\beta}-\beta^{\dagger}\right) \rightarrow
N\left( 0,\Psi\right) $ where $\Psi=\Omega\Sigma\Gamma_{\beta^{\dagger}}%
\Sigma$ with 
\begin{equation*}
\Gamma_{\beta^{\dagger}}=E_{\omega_{op}}\left[ VV^{T}\left\{ \sigma
^{2}+\left( \Pi_{\omega_{op}}\left[ \gamma\left( X\right) |\left[ V\right]
^{\perp}\right] \right) ^{2}b\left( X\right) \right\} \right] . 
\end{equation*}

b) $\sqrt{n}\left\{ \mathbb{P}_{n,\omega_{op}}\left( \widehat{\beta}%
^{T}V\right) -E_{\omega_{op}}\left( \beta^{\dagger T}V\right) \right\}
\rightarrow N\left( 0,\Lambda_{\omega_{op}}\right) $ where $\Lambda
_{\omega_{op}}=\Lambda_{\omega_{op},1}+\Lambda_{\omega_{op},2}+\Lambda
_{\omega_{op},3},$ 
\begin{equation*}
\Lambda_{\omega_{op},1}=\Omega\left[ \sigma^{2}+E_{\omega_{op}}\left[ \left(
\Pi_{\omega_{op}}\left[ \gamma\left( X\right) |\left[ V\right] ^{\perp}%
\right] \right) ^{2}b\left( X\right) \right] \right] 
\end{equation*}%
\begin{equation*}
\Lambda_{\omega_{op},2}=\Omega E_{\omega_{op}}\left[ \omega_{op}\left(
X\right) \left\{ V^{T}\beta^{\dagger}-E_{\omega_{op}}\left(
V^{T}\beta^{\dagger}\right) \right\} ^{2}\right] 
\end{equation*}
and%
\begin{equation*}
\Lambda_{\omega_{op},3}=2\Omega E_{\omega_{op}}\left[ \omega_{op}\left(
X\right) \Pi_{\omega_{op}}\left[ \gamma\left( X\right) |\left[ V\right]
^{\perp}\right] \left\{ V^{T}\beta^{\dagger}-E_{\omega_{op}}\left(
V^{T}\beta^{\dagger}\right) \right\} \right] 
\end{equation*}

c) $\sqrt{n}\left\{ \mathbb{P}_{n}\left( \widehat{\beta}^{T}V\right) -{\tau}%
_{pop}\right\} \rightarrow N\left( 0,\Lambda\right) $ where $%
\Lambda=\Lambda_{1}+\Lambda_{2}+\Lambda_{3},$ 
\begin{equation*}
\Lambda_{1}=\Omega^{-1}E_{\omega_{op}}\left[ \left\{ \Pi_{\omega_{op}}\left[
\omega_{op}\left( X\right) ^{-1}|\left[ V\right] \right] \right\}
^{2}\left\{ \sigma^{2}+\left( \Pi_{\omega_{op}}\left[ \gamma\left( X\right) |%
\left[ V\right] ^{\perp}\right] \right) ^{2}b\left( X\right) \right\} \right]
\end{equation*}%
\begin{equation*}
\Lambda_{2}=E\left[ \left\{ V^{T}\beta^{\dagger}-E_{\omega_{op}}\left(
V^{T}\beta^{\dagger}\right) \right\} ^{2}\right] 
\end{equation*}
and%
\begin{equation*}
\Lambda_{3}=2\Omega^{-1}E_{\omega_{op}}\left[ \Pi_{\omega_{op}}\left[
\omega_{op}\left( X\right) ^{-1}|\left[ V\right] \right] \Pi_{\omega _{op}}%
\left[ \gamma\left( X\right) |\left[ V\right] ^{\perp}\right] \left\{
V^{T}\beta^{\dagger}-E_{\omega_{op}}\left( V^{T}\beta^{\dagger }\right)
\right\} \right] 
\end{equation*}

d) If, in addition, model $\left( \ref{gamma}\right) $ for $\gamma\left(
X\right) $ holds, we have that $\Lambda_{\omega_{op},3}=\Lambda_{3}=0$, 
\begin{equation*}
\Lambda_{\omega_{op}}=\Omega\left\{ \sigma^{2}+E_{\omega_{op}}\left[
\omega_{op}\left( X\right) \left\{
V^{T}\beta^{\dagger}-E_{\omega_{op}}\left( V^{T}\beta^{\dagger}\right)
\right\} ^{2}\right] \right\} 
\end{equation*}
and%
\begin{equation*}
\Lambda=\Omega^{-1}\sigma^{2}E_{\omega_{op}}\left[ \left\{ \Pi_{\omega_{op}}%
\left[ \omega_{op}\left( X\right) ^{-1}|\left[ V\right] \right] \right\} ^{2}%
\right] +var\left\{ \gamma\left( X\right) \right\} 
\end{equation*}

Because under part d) of the preceding Theorem, model $\left( \ref{gamma}%
\right) $ for $\gamma\left( X\right) $ is correct, $\mathbb{P}_{n}\left( 
\widehat{\beta}^{T}V\right) $ is a consistent estimator of $\tau_{pop}.$ Of
course, under the conditions of part d), $\widehat{\tau}_{B-DR}\left( 
\widehat{m}_{REG},\widehat{\pi}\right) $ (or any other DR estimator of $%
\tau_{pop}$ in model $\mathcal{U}_{0}\left( \mathcal{F}\times\mathcal{P}%
\right) )$ is a CAN estimator of $\tau_{pop}$ with asymptotic variance $%
\left( \ref{vardr}\right) $ which can be re-written as $\Omega^{-1}%
\sigma^{2}E_{\omega_{op}}\left[ \left\{ \omega_{op}\left( X\right)
^{-1}\right\} ^{2}\right] +var\left\{ \gamma\left( X\right) \right\} .$
Thus, the saving in variance associated with assuming model $\left( \ref%
{gamma}\right) $ for $\gamma\left( X\right) $ is $\Omega^{-1}\sigma
^{2}E_{\omega_{op}}\left[ \left\{ \Pi_{\omega_{op}}\left[ \omega _{op}\left(
X\right) ^{-1}|\left[ V\right] ^{\perp}\right] \right\} ^{2}\right] .$ This
saving is 'for free' if indeed the null hypothesis $\left( \ref{null}\right) 
$ of no treatment effect at any level of $X$ holds; its practical benefit
being that Wald confidence intervals for $\tau_{pop}$ centered at $\mathbb{P}%
_{n}\left( \widehat{\beta}^{T}V\right) $ will tend to be shorter than those
centered at $\widehat{\tau}_{B-DR}$ when the null hypothesis $\left( \ref%
{null}\right) $ holds. Below, we report the results of a simulation
experiment that demonstrates this practical benefit.

The trade-off, of course, is that when $\left( \ref{gamma}\right) $ is
false, and thus $\gamma\left( x\right) $ is not a constant function and the
null-hypothesis $\left( \ref{null}\right) $ is false, $\mathbb{P}_{n}\left( 
\widehat{\beta}^{T}V\right) $ may be inconsistent for $\tau_{pop}$ while $%
\widehat{\tau}_{B-DR}\left( \widehat{m}_{REG},\widehat{\pi}\right) $ remains
consistent if either $\left( \ref{prop-mod}\right) $ holds or $d\neq k$ and $%
\left( \ref{total}\right) $ holds. However, as noted above, when $%
\gamma\left( x\right) $ is not a constant function, the estimators used in
current practice are also inconsistent for $\tau_{pop},$\ since they are
estimators of $\tau_{pop,\omega}$\ for $\omega\left( \cdot\right) $\
different from the identity. Obviously, one option for decreasing the bias
of $\mathbb{P}_{n}\left( \widehat{\beta}^{T}V\right) $ is to increase the
dimension $d$ of the vector $V.$ However, this decrease is at the cost of an
increase in variance when indeed the model $\left( \ref{gamma}\right) $ for $%
\gamma\left( x\right) $ has already been correct with the initial smaller
dimension. If the model $\left( \ref{gamma}\right) $ for $\gamma\left(
x\right) $ is already correct for some $d,$ then increasing the
dimensionality of the vector $v^{d}\left( x\right) $ has the effect of
increasing the estimator variance upwards towards the asymptotic variance of 
$\widehat{\tau}_{B-DR}\left( \widehat{m}_{REG},\widehat{\pi}\right) ,$ i.e.
the non-parametric variance bound.

The nature of the variance-bias trade-off of $\mathbb{P}_{n}\left( \widehat{%
\beta}^{T}V\right) $ as an estimator of $\tau_{pop}$ is best understood when
examining the formula for $\widehat{\beta}.$ Units outside the region of
overlap in estimated propensity scores (i.e., units with $T=1$ and $%
\pi\left( X;\widehat{\alpha}\right) $ near 1 and units with $T=0$ and $%
\pi\left( X;\widehat{\alpha}\right) $ near 0) will have a very small
contribution to the estimator $\widehat{\beta}$ because for them $%
T-\pi\left( X;\widehat{\alpha}\right) $ will be close to 0. Thus, the
contribution $\widehat{\beta}^{T}V_{i}$ to $\mathbb{P}_{n}\left( \widehat{%
\beta}^{T}V\right) $ of a unit $i$ outside the region of overlap, is the
result of an extrapolation; the contribution $\widehat{\beta}^{T}V_{i}$ is
the consequence of using the model $\beta^{T}v^{d}\left( x\right) $ to
approximate $\gamma\left( x\right) $ for $x$ outside the region of overlap
using mainly data from units within the region of overlap$.$ Such
extrapolation runs the risk of bias due to misspecification of the model for 
$\gamma\left( x\right) $ in exchange for confidence intervals for $%
\tau_{pop} $ that are typically narrow with near nominal finite sample
coverage when the model for $\gamma\left( x\right) $ is correct (which is
always the case under the null $\left( \ref{null}\right) $). Increasing the
dimension of $V$ and thus $\beta$ implies more parameters must be estimated
from the same data, which increases the contribution of units $i$ with $X_{i}
$ local (in Euclidean distance) to $x$ to the estimation of $\gamma\left(
x\right) ,$ effectively augmenting the previously small contribution of $%
T_{i}-\pi\left( X_{i};\widehat{\alpha}\right) $ of units $i$ outside the
region of overlap to the estimation of $\gamma\left( X_{i}\right) ;$ as a
consequence, model extrapolation (and bias) is reduced at the expense of an
increase in variance.

An alternative approach to decreasing extrapolation and large sample bias at
the expense of increased variance is by keeping the dimension of the model
for $\gamma\left( x\right) $ fixed but estimating $\beta$ with $\widehat{%
\beta }\left( c\right) $ using a function $c\left( x\right) $\ other than
the identity. As an extreme example, consider the estimator $\widehat{\beta }%
\left( c\right) $ that uses $c\left( X\right) =\widehat{\omega}_{op}\left(
X\right) ^{-1}=1/\left[ \pi\left( X;\widehat{\alpha}\right) \left\{
1-\pi\left( X;\widehat{\alpha}\right) \right\} \right] $. Then the effective
relative weight contributed by unit $i$ to the estimator $\widehat{\beta}%
\left( c\right) $ is proportional to $\left\{ T_{i}-\pi\left( X_{i};\widehat{%
\alpha}\right) \right\} \widehat{\omega}_{op}\left( X_{i}\right)
^{-1}=T_{i}/\pi\left( X_{i};\widehat{\alpha }\right) +$\ $\left(
1-T_{i}\right) /\left\{ 1-\pi\left( X_{i};\widehat{\alpha}\right) \right\} $%
. Thus subjects $i$\ outside the overlap region now contribute weights of
approximately one, which are no longer dramatically less than the weights
contributed by subjects with $\pi\left( X;\widehat{\alpha}\right) =1/2.\ $%
Thus the extrapolation involved in the estimate $\widehat{\beta}\left(
c\right) ^{T}V_{i}$\ is much decreased. However, the rare units with $%
\pi\left( X;\widehat{\alpha}\right) $\ near 0 but $T=1$\ or with $\pi\left(
X;\widehat{\alpha}\right) $\ near 1 but $T=0$\ now have enormous weight,
dwarfing the weight of approximately 1 they had when $c\left( \cdot\right) $%
\ was the identity. As a consequence, the variance of $\widehat{\beta}\left(
c\right) $\ for $c\left( X\right) =\widehat{\omega}_{op}\left( X\right)
^{-1} $\ can be very large compared to the case with $c\left( \cdot\right) $%
\ the identity. In fact, if the models for $\gamma\left( x\right) ,\pi\left(
x\right) ,$\ and $m_{0}\left( x\right) $\ are all correct, the asymptotic
distribution of $\mathbb{P}_{n}\left( \widehat{\beta}\left( c\right)
^{T}V\right) $\ with $c\left( X\right) =\widehat{\omega}_{op}\left( X\right)
^{-1}$\ is equal to that of $\widehat{\tau}_{B-DR}\left( \widehat{m}_{REG},%
\widehat{\pi}\right) $.

\subsection{Conditional versus unconditional Inference}

We can decompose the sources of variability of the limiting distribution of $%
\sqrt{n}\left( \widehat{\beta}-\beta^{\dagger}\right) $ when both models $%
\left( \ref{prop-mod}\right) $\ and $\left( \ref{reg-mod}\right) $ hold. To
do so, we write $\sqrt{n}\left( \widehat{\beta}-\beta^{\dagger}\right) =%
\sqrt{n}\left( \widehat{\beta}-\beta_{c}^{\dagger}\right) +\sqrt{n}\left(
\beta_{c}^{\dagger}-\beta^{\dagger}\right) $ where 
\begin{equation}
\beta_{c}^{\dagger}\equiv\mathbb{P}_{n,\omega_{op}}\left\{ \gamma
(X)V^{T}\right\} \mathbb{P}_{n,\omega_{op}}\left( VV^{T}\right) ^{-1}.
\label{betac}
\end{equation}
The quantity $\beta_{c}^{\dagger}$ is the $\omega_{op}-$ weighted least
squares coefficient in the regression of $\gamma\left( X_{i}\right) $ on $%
V_{i},i=1,...,n,$ and thus a function of $\mathbf{X}=\left\{
X_{1},...,X_{n}\right\} .$ It can be shown that when models $\left( \ref%
{prop-mod}\right) $\ and $\left( \ref{reg-mod}\right) $ hold, $\sqrt {n}%
\left( \widehat{\beta}-\beta_{c}^{\dagger}\right) $ and $\sqrt{n}\left(
\beta_{c}^{\dagger}-\beta^{\dagger}\right) $ are asymptotically independent
and converge to mean zero normal distributions with variances equal to $%
\Omega\Sigma\Gamma_{con}\Sigma$ and $\Omega\Sigma\Gamma_{\beta_{c}^{%
\dagger}}\Sigma$ where 
\begin{equation*}
\Gamma_{con}\equiv E_{\omega_{op}}\left[ VV^{T}\left\{ \sigma^{2}+\left\{
\Pi_{\omega_{op}}\left( \gamma\left( X\right) |\left[ V\right] ^{\perp
}\right) \right\} ^{2}\left\{ 1-2\pi\left( X\right) \right\} ^{2}\right\} %
\right] 
\end{equation*}
and 
\begin{equation*}
\Gamma_{\beta_{c}^{\dagger}}\equiv\Gamma_{\beta^{\dagger}}-\Gamma
_{con}=E_{\omega_{op}}\left[ \omega_{op}\left( X\right) VV^{T}\left\{
\Pi_{\omega_{op}}\left( \gamma\left( X\right) |\left[ V\right] ^{\perp
}\right) \right\} ^{2}\right] 
\end{equation*}
The convergence of $\sqrt{n}\left( \widehat{\beta}-\beta_{c}^{\dagger
}\right) $ to the $N\left( 0,\Omega\Sigma\Gamma_{con}\Sigma\right) $
distribution also holds conditionally on $\mathbf{X}$ when both models hold.
However, when model $\left( \ref{prop-mod}\right) $ is incorrect,
convergence of $\sqrt{n}\left( \widehat{\beta}-\beta_{c}^{\dagger}\right) $
to a mean zero normal distribution conditional on $\mathbf{X}$ requires the
validity of a different model than $\left( \ref{reg-mod}\right) ,$ namely
the model 
\begin{equation}
E\left\{ H\left( \beta_{c}^{\dagger}\right) |\mathbf{X}\right\}
=\theta_{c}^{\ast T}V^{\dagger}\text{ w.p.1.}  \label{newh-mod}
\end{equation}
which specifies that there exists $\theta_{c}^{\ast}\equiv\theta_{c}^{\ast
}\left( \mathbf{X}\right) $ such that with probability 1, for each $%
j=1,...,n,$ $E\left( Y_{j}|X_{j}\right) -\pi\left( X_{j}\right) \beta
_{c}^{\ast}\left( \mathbf{X}\right) ^{T}v^{d}\left( X_{j}\right)
=\theta_{c}^{\ast}\left( \mathbf{X}\right) v^{k}\left( X_{j}\right) .$\ 

\section{Simulation studies of estimators of the average treatment effect $%
\protect\tau_{pop}$.}

To compare the various candidate estimators of $\tau_{pop}$ described above,
we replicated a simulation experiment in Kang and Shafer (2007).
Specifically, we generated $n$ i.i.d. copies of $\left( Y,T,Z\right) $\ from 
$Z=\left( Z_{1},Z_{2},Z_{3},Z_{4}\right) ^{T}$ $\symbol{126}N\left(
0,I\right) ,$ $\pi\left( Z\right) \equiv P\left( T=1|Z\right) =$ expit$%
\left( -Z_{1}+0.5Z_{2}-0.25Z_{3}-0.1Z_{4}\right) ,$\newline
$Y=\gamma\left( Z\right)
T+210+27.4Z_{1}+13.7Z_{2}+13.7Z_{3}+13.7Z_{4}+\varepsilon,$ $\varepsilon%
\symbol{126}$N$\left( 0,1\right) $ with$\gamma\left( Z\right) \equiv0.$ We
can view the data as\ $n$ i.i.d. copies of an observed vector $O=\left(
T,Y,Z\right) $ with $T$ a binary treatment indicator, $Y$ a response, and $Z$
a covariate vector, satisfying the sharp null hypothesis $Y_{\left( 1\right)
}=Y_{\left( 0\right) }=Y$ w.p.1 and therefore the null hypothesis $\left( %
\ref{null}\right) $ as well as the assumptions of consistency, ignorability,
and strong positivity. Had we chosen treatment effect function $\gamma\left(
Z\right) $ non-zero, then $\left( \ref{null}\right) $ would not hold. Due to
space limitations, we only report simulation results under the null
hypothesis $\left( \ref{null}\right) .$

Following Kang and Shafer (2007)$,$ we defined the random vector $X=\left(
X_{1},X_{2},X_{3},X_{4}\right) ^{T}$ computed as $X=b\left( Z\right) $ where 
$b\left( z\right) =\left( b_{1}\left( z\right) ,b_{2}\left( z\right)
,b_{3}\left( z\right) ,b_{4}\left( z\right) \right) ^{T}$ satisfies $%
b_{1}\left( z\right) =\exp\left( z_{1}/2\right) ,$\newline
$b_{2}\left( z\right) =10+z_{2}/\left\{ 1+\exp\left( z_{1}\right) \right\}
,b_{3}\left( z\right) =\left( z_{1}z_{3}/25+0.6\right) ^{3}$ and $%
b_{4}\left( z\right) =\left( z_{2}+z_{4}+20\right) ^{2}.$ We note that the
transformed data $\left( Y,T,X\right) $ satisfies the semiparametric
regression model $\left( \ref{SRM}\right) $ with $\beta^{\ast}=0.$ It also
satisfies model $\left( \ref{total}\right) $ with $k=5,\nu^{k}\left(
x\right) =\left( 1,b^{-1}\left( x\right) \right) ^{T}=\left( 1,z\right) ^{T}$
where $b^{-1}\left( \cdot\right) $ is the inverse function of $b\left(
\cdot\right) .$

Results of our simulation study are given in table 1. The double-robust
difference estimators reported in rows 3-6 and 22-35 use correct outcome
model $\left( \ref{total}\right) $ with $k=5$ and $\nu^{k}\left( x\right)
=\left( 1,b^{-1}\left( x\right) \right) ^{T},$ i.e. for $t=0,1,$ each model
has an intercept and covariate vector $Z,$ while those reported in rows
11-14 and 17-20 use an incorrect outcome model $\left( \ref{total}\right) $
with $k=5$ and $\nu^{k}\left( x\right) =\left( 1,x\right) ^{T},$ i.e. with
intercept and covariate vector $X.$ Likewise, the propensity score model
assumed by the estimators in rows 1-6 and 17-20 is a correct logistic
regression model with intercept and covariate vector $Z\,\ $and that used by
the estimators in rows 9-14 and 22-35 is an incorrect logistic regression
models with an intercept but with the incorrect covariate vector $X.$

The estimators $P_{n}\left( \widehat{\beta}^{T}V\right) $ and $P_{n}\left( 
\widehat{\beta}_{OLS}^{T}V\right) $ reported in rows 7, 8 and 26 fit the $%
\gamma-$ model $\left( \ref{gamma}\right) $ and the working $H-$model $%
\left( \ref{reg-mod}\right) $ with $k=d=5$ and the transformation $\nu
^{k}\left( x\right) =\left( 1,b^{-1}\left( x\right) \right) ^{T}=\left(
1,z\right) ^{T}$ that makes both models $\left( \ref{gamma}\right) $ and $%
\left( \ref{reg-mod}\right) $ correct. The models reported in rows 15, 16
and 21 use $k=d=5$ but with the incorrect transformation $\nu^{k}\left(
x\right) =\left( 1,x\right) ^{T}.$ As a consequence the $H-$ model $\left( %
\ref{reg-mod}\right) $ is incorrect. However, the model $\left( \ref{gamma}%
\right) $ for $\gamma\left( x\right) $ is still correct, because the true
function $\gamma\left( x\right) $ is identically equal to 0 even under the
specification $\nu^{k}\left( x\right) =\left( 1,x\right) $\emph{.} The
working logistic regression models are those described above, their
correctness depending on the transformation used.

As predicted by theory, when the propensity score model is correct but the
outcome models $\left( \ref{total}\right) $ and the $H-$model $\left( \ref%
{reg-mod}\right) $ are incorrect, $\mathbb{P}_{n}\left( \widehat{\beta }%
^{T}V\right) $ greatly outperforms all other estimators. Specifically, $%
\mathbb{P}_{n}\left( \widehat{\beta}^{T}V\right) $ is nearly unbiased
[because the model for $\gamma\left( \cdot\right) \ $is correct] unlike $%
\mathbb{P}_{n}\left( \widehat{\beta}_{OLS}^{T}V\right) $ which is badly
biased in this circumstance. The asymptotically unbiased DR difference
estimators of $\tau_{pop}\ $of the form $\widehat{\tau}_{B-DR}\left( 
\widehat{m},\widehat{\pi}\right) $ have some small sample bias because $%
\pi\left( X\right) \left\{ 1-\pi\left( X\right) \right\} $ is extremely
variable; however the bias is still less than that of $\mathbb{P}_{n}\left( 
\widehat{\beta}_{OLS}^{T}V\right) $. Furthermore, $\mathbb{P}_{n}\left( 
\widehat{\beta}^{T}V\right) $ achieves huge savings in variance compared to
any of the estimators $\widehat{\tau}_{B-DR}\left( \widehat{m},\widehat{\pi }%
\right) $. Additionally, upon comparing rows 17 and 18, we see that $%
\widehat{\tau}_{B-DR}\left( \widehat{m}_{ITER-WLS},\widehat{\pi}%
_{ITER-WLS}\right) $ has a variance roughly 30 \% smaller than that of 
\newline
$\widehat{\tau}_{B-DR}\left( \widehat{m}_{WLS},\widehat{\pi}_{WLS}\right) ,$
thus confirming the theory of Section 3.1.

When the outcome model $\left( \ref{total}\right) $ and the $H-$models are
correct, the double-robust difference estimators in rows 22-25, and the
estimators $\mathbb{P}_{n}\left( \widehat{\beta}_{OLS}^{T}V\right) $ and $%
\mathbb{P}_{n}\left( \widehat{\beta}^{T}V\right) $ of rows 7 and 26
respectively, behave nearly identically, essentially because of the
exceedingly small error variance in the regression of $Y$ on $T$ and $Z.$

When both the outcome model and the propensity score model are wrong, the
estimator $\mathbb{P}_{n}\left( \widehat{\beta}^{T}V\right) $ has the
smallest bias and MSE, although its dominance over the other estimators is
less than when the propensity score model is correct.

\section{Model Selection and Sensitivity analysis}

Consider again the above simulation experiment in which the observed data
are $n$ i.i.d. copies of $\left( Y,T,X\right) $ and both the linear logistic
model for the propensity score and the linear outcome regression model $%
\left( \ref{total}\right) $ are wrong because they use the incorrect
transformation\emph{\ }$\nu^{k}\left( x\right) =\left( 1,x\right) ^{T}$. In such a setting RSGR argue that one often cannot be certain, owing to
lack of power, that a chosen model for the propensity score is nearly
correct, even if it passes a standard goodness of fit test. Therefore a
number of models for the propensity score with different subsets of the
covariates, different orders of interactions, and different dimensions of
the parameter vector should be fit to the data. Similarly a number of
different outcome models $\left( \ref{total}\right) $ should be fit. RSGR
raise the question: once fit, how should these many candidate models be used
in the estimation of the mean of $Y?$

RSGR mentioned that one approach is to follow the suggestion of van der Laan
(2005) and use $k-fold$ cross-validation (using squared error loss for the
outcome model $\left( \ref{total}\right) $ and minus log likelihood loss for
the linear logistic propensity models) to select the propensity and outcome
regression model that are to be used in the construction of a DR\ estimator
with the standard squared error loss function. In this section we report the
results of a simulation experiment in which we implement van der Laan's
suggestion and use 5 fold cross validation to choose the propensity and
outcome regression model. Due to space limitations, we only consider six
linear logistic propensity models $\left( \emph{\ref{prop-mod}}\right) $ and
four linear outcome regression models $\left( \ref{total}\right) $, and a
single DR difference estimator of $\tau_{pop}$ - the estimator $\widehat{%
\tau }_{B-DR}\left( \widehat{m}_{WLS},\widehat{\pi}\right) .$ Thus our
results should be viewed as illustrative, not definitive. The six propensity
and four outcome regression models are described in Table 2. None of the ten
models are correctly specified.

We compare the performance of the estimator $\widehat{\tau}_{B-DR}\left( 
\widehat{m}_{WLS},\widehat{\pi}\right) $ with the propensity and outcome
regression model selected by 5-fold cross validation with the results of a
novel alternative approach to model selection outlined by RSGR that
leverages the fact that the estimator $\widehat{\tau}_{B-DR}\left( \widehat{m%
}_{WLS},\widehat{\pi}\right) $ is double-robust.

Our alternative approach is as follows. Define $\widehat{\tau}_{ij}$ to be
the DR estimator $\widehat{\tau}_{B-DR}\left( \widehat{m}_{WLS},\widehat{\pi 
}\right) $ that uses the fitted values from the $i^{th}$ propensity model
and the $j^{th}$ outcome model\textbf{.} Now, if the $i^{th}$ propensity
model is correct, all four estimators in the set $\mathcal{E}%
_{p,i}\equiv\left\{ \widehat{\tau}_{ij};\text{ }j=1,....,4\right\} $ will be
CAN estimators of $\tau_{pop}.$ Thus, an $\alpha-$ level test of the
homogeneity hypothesis $H_{pi}:E^{A}\left( \widehat{\tau}_{i1}\right)
=E^{A}\left( \widehat{\tau }_{ij}\right) \ $for all $j\in\left\{
2,3,4\right\} $ (where $E^{A}\left( \cdot\right) $ stands for large sample
mean, i.e. the probability limit of, $\cdot)$ is also an $\alpha-$ level
goodness of fit test for the propensity model that is directly relevant to
its use in a double-robust estimator of $\tau_{pop}.$ Similarly if the $%
j^{th}$ outcome model is correct, all six estimators in the set $\mathcal{E}%
_{o,j}\equiv\left\{ \widehat{\tau}_{ij};\text{ }i=1,....,6\right\} $ will be
CAN for $\tau_{pop}\ $and an $\alpha-$ level test of the homogeneity
hypothesis $H_{oj}:E^{A}\left( \widehat{\tau}_{1j}\right) =E^{A}\left( 
\widehat{\tau}_{ij}\right) \ $for all $i\in\left\{ 2,...,6\right\} $ is an $%
\alpha-$ level test of goodness of fit for the outcome model. This suggests
that one choose as a final estimator of $\tau_{pop}$ the DR\ estimator $%
\widehat{\tau}_{i^{\ast}j^{\ast}},$ where $i^{\ast}$ is the $i$ for which
the test of the hypothesis $H_{pi}$ gave the largest p-value and $j^{\ast}$
is the $j$ for which the test of the hypothesis $H_{oj}$ gave the largest
p-value. However, this method of selecting $i^{\ast }$ and $j^{\ast}$ may be
non-optimal for two reasons. First, it could easily select a misspecified
propensity model $i$ for which the power of the test of the hypothesis $%
H_{pi}$ is particularly poor and similarly for the outcome regression. This
remark implies that some measure of the spread of the elements of $\mathcal{E%
}_{p,i}$ and $\mathcal{E}_{o,j}$ should also contribute to the selection $%
i^{\ast}$ and $j^{\ast}.$ Therefore we also consider choosing the DR\
estimator $\widehat{\tau}_{i^{\ast}j^{\ast}}$ with $i^{\ast}$ now the $i$
for which the empirical standard deviation of the $\widehat{\tau }_{ij};$ $%
j=1,....,4$ is the least and $j^{\ast}$ now the $j$ for which the empirical
standard deviation of the $\widehat{\tau}_{ij};$ $i=1,....,6\ $is the least.
Further to possibly increase sensitivity, we also consider replacing the
standard deviation by the range in the previous definition of $i^{\ast}$ and 
$j^{\ast}.$ None of the above alternative methods exploit the fact that if $%
i^{\ast}$ and $j^{\ast}$ are correct then $E^{A}\left( \widehat{\tau}%
_{ij^{\ast}}\right) =E^{A}\left( \widehat{\tau}_{i^{\ast}j}\right) $ for all 
$i$ and $j,$ suggesting that an optimal method should perhaps select $%
i^{\ast}$ and $j^{\ast}$ jointly. Therefore we also considered selecting the
DR\ estimator $\widehat{\tau}_{i^{\ast}j^{\ast}}$ with $\left(
i^{\ast},j^{\ast}\right) $ now the pair $\left( i,j\right) $ that minimizes
the sum of: (i) the empirical standard deviation of the $\widehat{\tau}%
_{ij}; $ $j=1,2,3,4,$ (ii) the empirical standard deviation of the $\widehat{%
\tau}_{ij};$ $i=1,....,6$ , and (iii) a positive constant $c$ times $%
\left\vert \widehat{\tau}_{i+,av}-\widehat{\tau}_{+j,av}\right\vert $ where $%
\widehat{\tau}_{i+,av}=\dsum \limits_{j=1}^{4}\widehat{\tau}_{ij}/4$ and $%
\widehat{\tau}_{+j,av}=\dsum \limits_{j=1}^{6}\widehat{\tau}_{ij}/6$. Since
we have yet to develop a theory by which to choose $c$, in our simulation
study we simply tried the choices $c=1,2,3,4$ . Note the choice $c=0$ is
equivalent to the separate selection of $i^{\ast}$ and $j^{\ast}$ based on
the standard deviation.

Interestingly, a by-product of this model selection algorithm is that the
matrix with elements $\widehat{\tau}_{ij}$ immediately provides an informal
sensitivity analysis; we directly observe the sensitivity of our DR
estimator to the choice of propensity and outcome regression model.

Table 3 reports the results for each of the 24 (non-data adaptive)
estimators $\widehat{\tau}_{ij}$, $i=1,....,6,j=1,2,3,4.$ We see that $%
\widehat{\tau }_{2,4}$ has the smallest Monte Carlo (MC) MSE of 7.908 among
all of them. The results for the various data-adaptive methods for model
selection described above are given in Table 4 as well as the results for
the oracle estimator. The oracle estimator is defined as the $\widehat{\tau}%
_{ij}$ that is closest to the true $\tau_{pop}.$ It is an infeasible
estimator which serves as an unobtainable benchmark for comparison of the
performance of the feasible procedures. The oracle estimator, of course,
need not be the same $\widehat{\tau}_{ij}$ in all data realizations. Thus,
it is guaranteed to have MSE at least as small as the smallest MSE of the 24
non-data adaptive estimators $\widehat{\tau}_{ij},$ (i.e. the MSE of $%
\widehat{\tau}_{2,4}$ in our simulations). Equality between the MC MSE's of
the oracle and the best non-data adaptive estimator would hold only if the
best non-data adaptive estimator happened to be the estimator closest to $%
\tau_{pop}$ in each single data realization of the simulation study, an
unlikely situation that did not occur in our simulations. Indeed, the MC MSE
of the oracle estimator was 5.824, roughly 25\% smaller than the MC MSE of $%
\widehat{\tau}_{2,4}$.

From Table 4, we see that the MC MSE of data-adaptive estimators based on
the range and the standard deviation were essentially equal and were better
than the MC MSE of the 5-fold cross validation estimator, of any
joint-selection estimator regardless of the value of $c$, and much better
than the MSE of the estimator based on the Wald test of homogeneity. In
fact, the data-adaptive estimators based on the range and the standard
deviation had smaller MC MSE than 22 of the 24 non-data adaptive estimators
estimators in Table 3, which is a rather gratifying result. In contrast the
estimator based on the Wald test of homogeneity has a MC MSE worse than all
non-data adaptive estimators but the estimator $\widehat{\tau}_{1,1}$, the
estimator based on the most parsimonious model for both the propensity score
and the outcome regression. This result is consistent with our concern,
expressed above, that due to lack of power, the models selected by the Wald
test would be too parsimonious. In summary, based on the results of this one
simulation, our novel approach to model selection based on either the range
or the standard deviation appears promising, although additional
investigation, both theoretical and by simulation, is necessary before any
general recommendation could be made. Note that although we applied our
novel model selection approach to the DR\ estimator $\widehat{\tau}%
_{B-DR}\left( \widehat{m}_{WLS},\widehat{\pi }\right) $ of the parameter $%
\tau_{pop},$ the approach is applicable to the estimation of any parameter
that admits a DR estimator. Robins and Rotnitzky (2001) provide methods for
determining whether a particular parameter of interest admits a DR estimator.

\section{Appendix.}

\textbf{A.1.} \textbf{Sketch of the} \textbf{derivation of the asymptotic
variance of} $\widehat{\tau}_{B-DR}$ \textbf{under model }$\mathcal{U}%
_{0}\left( \mathcal{F}\times\mathcal{P}\right) $\textbf{\ when both models }$%
\left( \ref{prop-mod}\right) $\textbf{\ and }$\left( \ref{total}\right) $%
\textbf{\ hold.}

When models $\left( \ref{prop-mod}\right) $\textbf{\ }and\textbf{\ }$\left( %
\ref{total}\right) $ hold, $\widehat{\alpha},\widehat{\beta}_{OLS}$ and $%
\widehat{\theta}_{OLS}$ converge in probability to $\alpha^{\ast},\beta
^{\ast}$ and $\theta^{\ast}.$ Furthermore, $\beta^{\ast T}V=\gamma\left(
X\right) $ and $\theta^{\ast}V^{\dagger}=m_{0}\left( X\right) .$ We have 
\begin{align}
\sqrt{n}\left( \widehat{\tau}_{B-DR}-\tau_{pop}\right) & =\sqrt{n}\left( 
\widehat{\mu}_{1,B-DR}-\mu_{1}\right) -\sqrt{n}\left( \widehat{\mu}%
_{0,B-DR}-\mu_{0}\right)  \notag \\
& =\sqrt{n}\left[ \mathbb{P}_{n}\left[ \frac{T}{\pi\left( X\right) }\left\{
Y-E\left( Y|T,X\right) \right\} \right] +\mathbb{P}_{n}\left\{ \gamma\left(
X\right) +m_{0}\left( X\right) \right\} -\mu_{1}\right]  \notag \\
& -\sqrt{n}\left[ \mathbb{P}_{n}\left[ \frac{\left( 1-T\right) }{1-\pi\left(
X\right) }\left\{ Y-E\left( Y|T,X\right) \right\} \right] +\mathbb{P}%
_{n}\left\{ \gamma\left( X\right) \right\} -\mu_{0}\right] +o_{p}\left(
1\right)  \notag \\
& =\sqrt{n}\mathbb{P}_{n}\left[ \frac{T-\pi\left( X\right) }{\omega
_{op}\left( X\right) }\left\{ Y-E\left( Y|T,X\right) \right\} +\gamma\left(
X\right) -\tau_{pop}\right] +o_{p}\left( 1\right)  \label{A1}
\end{align}
where the second equality is obtained after standard Taylor expansions and
noticing that $\left. \frac{\partial\widehat{\mu}_{j,B-DR}}{\partial\alpha }%
\right\vert _{\alpha^{\ast}}=\left. \frac{\partial\widehat{\mu}_{j,B-DR}}{%
\partial\beta}\right\vert _{\beta^{\ast}}=\left. \frac{\partial\widehat {\mu}%
_{j,B-DR}}{\partial\theta}\right\vert _{\theta^{\ast}}=o_{p}\left( 1\right)
. $

Now, $\left( \ref{A1}\right) $ and the Central Limit Theorem give that 
\begin{align*}
var^{A}\left\{ \sqrt{n}\left( \widehat{\tau}_{B-DR}-\tau_{pop}\right)
\right\} & =var\left\{ \frac{T-\pi\left( X\right) }{\omega_{op}\left(
X\right) }\left\{ Y-E\left( Y|T,X\right) \right\} +\gamma\left( X\right)
-\tau_{pop}\right\} \\
& =\sigma^{2}E\left[ \frac{\left\{ T-\pi\left( X\right) \right\} ^{2}}{%
\omega_{op}\left( X\right) }\right] +var\left\{ \gamma\left( X\right)
\right\} \\
& =\sigma^{2}E\left\{ \omega_{op}\left( X\right) ^{-1}\right\} +var\left\{
\gamma\left( X\right) \right\}
\end{align*}

The formula for $var^{A}\left\{ \sqrt{n}\left( \widehat{\tau}%
_{B-DR}-\tau_{con}\right) \right\} $ follows from the Central Limit Theorem
after noticing that 
\begin{align*}
\sqrt{n}\left( \widehat{\tau}_{B-DR}-\tau_{con}\right) & =\sqrt{n}\left( 
\widehat{\tau}_{B-DR}-\tau_{pop}\right) +\sqrt{n}\left( \tau_{pop}-\mathbb{P}%
_{n}\left\{ \gamma\left( X\right) \right\} \right) \\
& =\sqrt{n}\mathbb{P}_{n}\left[ \frac{T-\pi\left( X\right) }{\omega
_{op}\left( X\right) }\left\{ Y-E\left( Y|T,X\right) \right\} \right]
+o_{p}\left( 1\right)
\end{align*}
where the second identity follows from $\left( \ref{A1}\right) .$

\textbf{A.2} \textbf{Sketch of the proof that }$\widehat{\mu}_{t,B-DR}\left( 
\widehat{m}_{t},\widehat{\pi}\right) $\textbf{\ is at least as efficient as }%
$\widehat{\mu}_{t,IPW}$\textbf{\ when the propensity and outcome model are
correct but not necessarily if the outcome model is incorrect.}

Quite generally, suppose we specify a logistic regression model $\pi\left(
x\right) =\pi_{gen}\left( x;\widehat{b},\eta^{\ast}\right) $ where $%
\eta^{\ast}$ is an unknown parameter vector $\widehat{b}\left( x\right)
=\left( b_{1}\left( \widehat{\delta};x\right) ,...,b_{s}\left( \widehat{%
\delta};x\right) \right) ^{T},$ $b_{j}\left( \delta;x\right) $ is a smooth
function of a parameter vector $\delta$ for each $x,j=1,...,s,$ and $%
\widehat{\delta}=\delta^{\ast}+O_{p}\left( n^{-1/2}\right) ,$ and where for
any conformable vector function $u\left( x\right) $ 
\begin{equation}
\pi_{gen}\left( x;u,\eta\right) \equiv\text{expit}\left\{ \eta^{T}u\left(
x\right) \right\}  \label{prop-gen}
\end{equation}
Let $\widehat{\eta}$ be the ML estimator of $\eta^{\ast}$ under model $%
\left( \ref{prop-gen}\right) $ solving the score equations $\mathbb{P}%
_{n}\left\{ S_{\widehat{b}}\left( \eta\right) \right\} =0$ where for any
conformable function $u\left( x\right) ,$ $S_{u}\left( \eta\right) \equiv
u\left( X\right) \left[ T-\pi_{gen}\left( X;u,\eta\right) \right] $ and let $%
\widehat{\pi}_{gen}\left( X\right) =$ $\pi_{gen}\left( x;\widehat {b},%
\widehat{\eta}\right) $. An easy calculation shows that $\widehat{\mu }%
_{t,B-DR}\left( \widehat{m}_{t},\widehat{\pi}_{gen}\right) $ satisfies 
\newline
$\mathbb{P}_{n}\left\{ U_{t}\left( \widehat{\mu}_{t,B-DR}\left( \widehat{m}%
_{t},\widehat{\pi}_{gen}\right) ,\widehat{\pi}_{gen},\widehat {m}_{t},%
\widehat{\mu}_{t}\right) \right\} =0$ where $\widehat{\mu}_{t}=\mathbb{P}%
_{n}\left\{ \widehat{m}_{t}\left( X\right) \right\} $ and for any constants $%
a$ and $c,$ and any functions $p\left( x\right) $ and $g\left( x\right) ,$%
\begin{equation*}
U_{t}\left( a,p,g,c\right) =\frac{I\left( T=t\right) }{p_{t}\left( X\right) }%
\left( Y-a\right) -\left\{ \frac{I\left( T=t\right) }{p_{t}\left( X\right) }%
-1\right\} \left\{ g\left( X\right) -c\right\} 
\end{equation*}
with $p_{t}\left( x\right) =p\left( x\right) ^{t}\left( 1-p\left( x\right)
\right) ^{1-t}.$ Let $b^{\ast}\left( x\right) =\left( b_{1}\left(
\delta^{\ast};x\right) ,...,b_{s}\left( \delta^{\ast};x\right) \right) ^{T}.$
Standard Taylor expansion arguments imply that when the model $\pi\left(
x\right) =\pi_{gen}\left( x;b^{\ast},\eta^{\ast}\right) $ is correct, 
$\sqrt{n}\left\{ \widehat{\mu}_{t,B-DR}\left( \widehat {m}_{t},\widehat{\pi}%
_{gen}\right) -\mu_{t}\right\} $ converges to a mean zero normal
distribution with variance equal to the variance of 
\begin{equation}
U_{t}\ \left( \mu_{t},\pi,m_{t}^{\ast},\mu_{t}^{\ast}\right) -E\left\{
U_{t}\left( \mu_{t},\pi,m_{t}^{\ast},\mu_{t}^{\ast}\right)
S_{b^{\ast}}\right\} var\left( S_{b^{\ast}}\right) ^{-1}S_{b^{\ast}}
\label{qq1}
\end{equation}
where $S_{b^{\ast}}=S_{b^{\ast}}\left( \eta^{\ast}\right) $ and $%
m_{t}^{\ast}\left( x\right) $ and $\mu_{t}^{\ast}$ are the probability
limits of $\ \widehat{m}_{t}\left( x\right) $ and $\mathbb{P}_{n}\left\{ 
\widehat {m}_{t}\left( X\right) \right\} $ respectively$.$ Analogous
arguments show that $\sqrt{n}\left( \widehat{\mu}_{t,IPW}-\mu_{t}\right) $
converges to a mean zero normal distribution with variance equal to 
\begin{equation}
I\left( T=t\right) \left( Y-\mu_{t}\right) /\pi_{t}\left( X\right) -E\left[
\left\{ I\left( T=t\right) \left( Y-\mu_{t}\right) /\pi _{t}\left( X\right)
\right\} S_{b^{\ast}}\right] var\left( S_{b^{\ast}}\right) ^{-1}S_{b^{\ast}}
\label{qq2}
\end{equation}
and $\sqrt{n}\left( \widehat{\mu}_{t,IPW}-\mu_{t}\right) $ converges to a
mean zero normal distribution with variance equal to 
\begin{equation}
I\left( T=t\right) \left( Y-\mu_{t}\right) /\pi_{t}\left( X\right)
+\mu_{t}S_{g_{t,2}}\left( X\right) -E\left[ \left\{ I\left( T=t\right)
\left( Y-\mu_{t}\right) /\pi_{t}\left( X\right) +\mu_{t}S_{g_{t,2}}\left(
X\right) \right\} S_{b^{\ast}}\right] var\left( S_{b^{\ast}}\right)
^{-1}S_{b^{\ast}}  \label{qq3}
\end{equation}
where $S_{g_{t,2}}\left( x\right) =1/\pi\left( x\right) .$ The quantities $%
\left( \ref{qq1}\right) ,$ $\left( \ref{qq2}\right) $ and $\left( \ref{qq3}%
\right) $ are the residuals from the population least squares of the
quantities $U\left( \mu_{t},\pi,m_{t}^{\ast},\mu_{t}^{\ast}\right) $, $%
I\left( T=t\right) \left( Y-\mu_{t}\right) /\pi_{t}\left( X\right) ,$ and 
$I\left( T=t\right) \left( Y-\mu_{t}\right) /\pi_{t}\left( X\right)
+\mu_{t}S_{g_{t,2}}\left( X\right) $, respectively, on $S_{b^{\ast}}$.
Neither $\left( \ref{qq2}\right) $ nor $\left( \ref{qq3}\right) $ is
guaranteed to have variance either smaller or larger than the variance of $%
\left( \ref{qq1}\right) $ and consequently, no general statement can be made
about the asymptotic relative efficiency of $\widehat{\mu}_{t,B-DR}\left( 
\widehat{m}_{t},\widehat{\pi}_{gen}\right) $ with respect to either $%
\widehat{\mu}_{t,IPW}$ \ or $\widehat{\mu}_{t,HT}.$ However, in the special
case in which the outcome model $\left( \ref{total}\right) $ is also
correctly specified, it holds that $m_{t}^{\ast }\left( X\right) $ is equal
to the true conditional mean function $m_{t}\left( X\right) \equiv E\left(
Y|T=t,X\right) $ and $\mu_{t}^{\ast }=\mu_{t}.$ A little algebra then shows
that $U_{t}\left( \mu_{t},\pi ,m_{t}^{\ast},\mu_{t}^{\ast}\right)
=U_{t}\left( \mu_{t},\pi,m_{t},\mu _{t}\right) $ is equal to 
\begin{equation*}
I\left( T=t\right) \left( Y-\mu_{t}\right) /\pi_{t}\left( X\right) -\left\{
I\left( T=t\right) -\pi_{t}\left( X\right) \right\} E\left[ I\left(
T=t\right) \left( Y-\mu_{t}\right) /\pi_{t}\left( X\right) |X\right] 
\end{equation*}
which is orthogonal to, i.e. uncorrelated with, $S_{u}$ for any function $%
u\left( x\right) $ and in particular with $S_{b^{\ast}}$. Consequently, in
such case, $\left( \ref{qq1}\right) $ is equal to $U_{t}\left(
\mu_{t},\pi,m_{t},\mu_{t}\right) $ and therefore $var^{A}\left\{ \widehat{%
\mu }_{t,B-DR}\left( \widehat{m}_{t},\widehat{\pi}\right) \right\} $ is
equal to $var\left\{ U_{t}\left( \mu_{t},\pi,m_{t},\mu_{t}\right) \right\} .$
But $var\left\{ U_{t}\left( \mu_{t},\pi,m_{t},\mu_{t}\right) \right\} $ is
no greater than the variance of $\left( \ref{qq2}\right) ,$ and then no
greater than $var^{A}\left\{ \widehat{\mu}_{t,IPW}\right\} ,$ because $%
U_{t}\left( \mu_{t},\pi,m_{t},\mu_{t}\right) $ is the residual from
projection of on the space spanned by the set $\left\{ S_{u}:u\left(
x\right) \text{ arbitrary}\right\} $ while $\left( \ref{qq2}\right) $ is the
projection of the same random variable on the smaller space spanned by just
the component of $S_{b^{\ast}}.$ Also, $var\left\{ U_{t}\left( \mu
_{t},\pi,m_{t},\mu_{t}\right) \right\} $ is no greater than the variance $%
\left( \ref{qq3}\right) ,$ and therefore than $var^{A}\left\{ \widehat {\mu}%
_{t,IPW}\right\} ,$ because $\left( \ref{qq3}\right) $ is of the form $%
I\left( T=t\right) \left( Y-\mu_{t}\right) /\pi_{t}\left( X\right)
+S_{u^{\ast}}\left( X\right) $ for some $u^{\ast}\left( X\right) $ since $%
I\left( T=t\right) -\pi_{t}\left( X\right) =\left( -1\right) ^{t}\left\{
T-\pi\left( X\right) \right\} .$

\textbf{A.3 Sketch of the proof that }$\widehat{\mu}_{t,B-DR}\left( \widehat{%
m}_{t,ITER-WLS},\widehat{\pi}_{ITER-WLS}^{\left( t\right) }\right) $ and $\widehat{\mu}_{t,B-DR}\left( \widehat{m}_{t,REG},\widehat{\pi}%
_{ITER-REG}^{\left( t\right) }\right) $ \textbf{satisfy the properties i)
and ii) of Section 3.1. }

We show that properties i) and ii) hold for $\widehat{\mu}_{t,B-DR}\left( 
\widehat{m}_{t,ITER-WLS},\widehat{\pi}_{ITER-WLS}^{\left( t\right) }\right)
. $ The proof that these properties hold for $\widehat{\mu}_{t,B-DR}\left( 
\widehat{m}_{t,REG},\widehat{\pi}_{ITER-REG}^{\left( t\right) }\right) $
follows verbatim keeping only the assertions about the limiting distribution
of $\widehat{\mu}_{t,B-DR}\left( \widehat{m}_{t,WLS},\widehat{\pi}%
_{ITER-WLS}^{\left( t\right) }\right) ,$ i.e. disregarding the assertions
about $$\widehat{\mu}_{t,B-DR}\left( \widehat{m}_{t,ITER-WLS},\widehat{\pi }%
_{ITER-WLS}^{\left( t\right) }\right) ,$$ and replacing every appearance of
the symbol $WLS$ by the symbol $REG.$

For each $t=0,1,$ model $\pi\left( x\right) =\pi_{ITER-WLS}^{\left( t\right)
}\left( x;\alpha_{t}^{\ast},\varphi_{t,1}^{\ast},\varphi
_{t,2}^{\ast}\right) $ is of the form $\pi\left( x\right) =\pi_{gen}\left( x;%
\widehat{b},\eta^{\ast}\right) $ with $\eta^{\ast}=\left( \alpha_{t}^{\ast
T},\varphi_{t,1}^{\ast},\varphi_{t,2}^{\ast}\right) $ and $\widehat{b}\left(
X\right) =\left( q\left( X\right) ^{T},\widehat{g}_{t,1}\left( X\right) ,%
\widehat{g}_{t,2}\left( X\right) \right) ^{T}.$ When model $\pi\left(
x\right) =\pi\left( x;\alpha^{\ast}\right) ,$ with $\pi\left(
x;\alpha\right) $ as in $\left( \ref{st-prop}\right) ,$ is correct so is
model $\pi\left( x\right) = \pi_{ITER-WLS}^{\left( t\right) }\left(
x;\alpha_{t}^{\ast},\varphi_{t,1}^{\ast},\varphi_{t,2}^{\ast}\right) $ with $%
\varphi_{t,1}^{\ast}=\varphi_{t,2}^{\ast}=0$. Consequently, the probability
limits of $\widehat{\pi}_{ITER-WLS}^{\left( t\right) }\left( x\right) $ and $%
\widehat{\pi}\left( x\right) $ \emph{are equal} and so are the probability
limits of $\ \widehat{m}_{t,ITER-WLS}\left( x\right) $ and $\widehat{m}%
_{t,WLS}\left( x\right) $. It follows that $$\sqrt{n}\left\{ \widehat{\mu}%
_{t,B-DR}\left( \widehat{m}_{t,WLS},\widehat{\pi}_{ITER-WLS}^{\left(
t\right) }\right) -\mu_{t}\right\}$$ and
$$\sqrt {n}\left\{ \widehat{\mu}_{t,B-DR}\left( \widehat{m}_{t,ITER-WLS},%
\widehat{\pi}_{ITER-WLS}^{\left( t\right) }\right) -\mu_{t}\right\}$$
converge to the same normal limiting distribution with mean zero and
variance equal to the variance of $\left( \ref{qq1}\right) $ where $%
b^{\ast}\left( x\right) =\left( q\left( x\right) ^{T},g_{t,1}\left( x\right)
,g_{t,2}\left( x\right) \right) ^{T},$ $g_{t,1}\left( x\right)
=m_{t}^{\ast}\left( x\right) /\pi_{t}\left( x\right) ,$ $g_{t,2}\left(
x\right) =1/\pi_{t}\left( x\right) $, $m_{t}^{\ast}\left( x\right) $ is the
common probability limit of $\widehat{m}_{t,WLS}\left( x\right) $ and $%
\widehat{m}_{t,ITER-WLS}\left( x\right) ,$ and $\mu_{t}^{\ast}$ is the
probability limit of $\mathbb{P}_{n}\left\{ m_{t}^{\ast}\left( X\right)
\right\} .$ With these definitions $U_{t}\left(
\mu_{t},\pi,m_{t}^{\ast},\mu_{t}^{\ast}\right) $ is exactly equal to $%
I\left( T=t\right) \left( Y-\mu_{t}\right) /\pi_{t}\left( X\right)
-S_{g_{t,1}}+\mu_{t}^{\ast }S_{g_{t,2}}.$ This implies that $\left( \ref{qq1}%
\right) $ simplifies to $\left( \ref{qq2}\right) $ (with $b^{\ast}\left(
X\right) =\left( q\left( X\right) ^{T},g_{t,1}\left( X\right) ,g_{t,2}\left(
X\right) \right) ^{T}),$ because for $j=1,2,S_{g_{t,j}}-E\left[
S_{g_{t,j}}S_{b^{\ast}}^{T}\right] var\left( S_{b^{\ast}}\right)
^{-1}S_{b^{\ast}},$ being the residual from the population regression of $%
S_{g_{t,j}}$ on $S_{b^{\ast}}=\left(
S_{q}^{T},S_{g_{t,1}},S_{g_{t,2}}\right) ^{T},$ is equal to 0. The
asymptotic variance of $\widehat{\mu}_{t,IPW}$ is also equal to the variance
of expression $\left( \ref{qq2}\right) $ but with $b^{\ast }\left( X\right) $
redefined as $q\left( X\right) .$ We then conclude that $Var^{A}\left\{ 
\widehat{\mu}_{t,IPW}\right\} $ is greater than or equal to $Var^{A}\left\{ 
\widehat{\mu}_{t,B-DR}\left( \widehat{m}_{t,ITER-WLS},\widehat{\pi}%
_{ITER-WLS}^{\left( t\right) }\right) \right\} $ because the latter is equal
to the residual variance of the regression of $I\left( T=t\right) \left(
Y-\mu_{t}\right) /\pi_{t}\left( X\right) $ on $\left(
S_{q}^{T},S_{g_{t,1}},S_{g_{t,2}}\right) ^{T}$ and the former is equal to
the residual variance of the regression of $I\left( T=t\right) \left( Y-\mu
_{t}\right) /\pi_{t}\left( X\right) $ on $S_{q}$ alone. To show that $$
Var^{A}\left\{ \widehat{\mu}_{t,B-DR}\left( \widehat{m}_{t,ITER-WLS},%
\widehat{\pi}_{ITER-WLS}^{\left( t\right) }\right) \right\}$$ is no greater
than $Var^{A}\left\{ \widehat{\mu}_{t,HT}\right\} $ we note that $%
Var^{A}\left\{ \widehat{\mu}_{t,HT}\right\} $ is the variance of the
residual of $I\left( T=t\right) \left( Y-\mu_{t}\right) /\pi _{t}\left( X\right)
+\mu_{t}S_{g_{t,2}}$ on $S_{q}$ and that this is greater than or equal the
variance of the residual of $I\left( T=t\right) \left( Y-\mu_{t}\right)
/\pi_{t}\left( X\right) +\mu_{t}S_{g_{t,2}}$ on $\left(
S_{q}^{T},S_{g_{t,1}},S_{g_{t,2}}\right) ^{T}.$ However, this last variance
is equal to the variance of $\left( \ref{qq1}\right) $ with $b^{\ast}\left(
x\right) =\left( q\left( x\right) ^{T},g_{t,1}\left( x\right) ,g_{t,2}\left(
x\right) \right) ^{T},$ and hence equal to $Var^{A}\left\{ \widehat{\mu}%
_{t,B-DR}\left( \widehat{m}_{t,ITER-WLS},\widehat{\pi }_{ITER-WLS}^{\left(
t\right) }\right) \right\} ,$ because the residual of $\mu_{t}S_{g_{t,2}}$
on $\left( S_{q}^{T},S_{g_{t,1}},S_{g_{t,2}}\right) ^{T}$ is 0. Finally, $%
Var^{A}\left\{ \widehat{\mu}_{t,B-DR}\left( \widehat{m}_{t,WLS},\widehat{\pi}%
\right) \right\} $ is also greater than or equal to $Var^{A}\left\{ \widehat{\mu}_{t,B-DR}\left( \widehat {m}_{t,ITER-WLS},%
\widehat{\pi}_{ITER-WLS}^{\left( t\right) }\right) \right\} $ because the
former is the residual variance of the regression of $U_{t}\left(
\mu_{t},\pi,m_{t}^{\ast},\mu_{t}^{\ast}\right) $ on $S_{q}$ while the latter
is the residual variance of $U_{t}\left( \mu_{t},\pi
,m_{t}^{\ast},\mu_{t}^{\ast}\right) $ on $\left(
S_{q}^{T},S_{g_{t,1}},S_{g_{t,2}}\right) ^{T}$.

\textbf{A.4.} \textbf{Proof of the double-robustness of\ }$\widehat{\beta}$ 
\textbf{as an estimator of} $\beta^{\dagger}$ \textbf{in model} $\mathcal{U}%
_{1}\left( \mathcal{F}\times\mathcal{P}\right) $

The form of $\widehat{\beta}$ in $\left( \ref{betahat}\right) $ and the Law
of Large Numbers imply that $\widehat{\beta}$ converges in probability to $%
\beta^{\dagger\dagger}$ solving 
\begin{equation}
E\left[ V\left\{ T-\pi\left( X;\alpha^{\dagger}\right) \right\} \left\{
H\left( \beta\right) -\Pi\left[ E\left\{ H\left( \beta\right) |X\right\}
|V^{\dagger}\right] \right\} \right] =0  \label{main1}
\end{equation}
A Taylor expansion and the Central Limit Theorem imply that $\sqrt{n}\left( 
\widehat{\beta}-\beta^{\dagger\dagger}\right) $ converges in law to a normal
distribution. Since the solution of $\left( \ref{main1}\right) $ is unique,
it suffices to check that $\beta^{\dagger}$ solves $\left( \ref{main1}%
\right) $ whenever either model $\left( \ref{prop-mod}\right) $ or model $%
\left( \ref{reg-mod}\right) $ is correct. To do so, it is enough to show
that when either model is correct, 
\begin{equation}
E\left[ V\left\{ T-\pi\left( X;\alpha^{\dagger}\right) \right\} \left\{
H\left( \beta^{\dagger}\right) -\Pi\left[ E\left\{ H\left( \beta
^{\dagger}\right) |X\right\} |V^{\dagger}\right] \right\} \right] =E\left[
V\left\{ T-E\left( T|X\right) \right\} T\left\{ \gamma\left( X\right)
-V^{T}\beta^{\dagger}\right\} \right]  \label{main2}
\end{equation}
because the right hand side of $\left( \ref{main2}\right) $ is equal to $E%
\left[ V\omega_{op}\left( X\right) \left\{ \gamma\left( X\right)
-V^{T}\beta^{\dagger}\right\} \right] $ (which can be verified taking
conditional expectations with respect to $X$ inside the right hand side of $%
\left( \ref{main2}\right) ),$ and this in turn, is equal to 0, by definition
of $\beta^{\dagger}.$

That $\left( \ref{main2}\right) $ holds when $\left( \ref{prop-mod}\right) $
is correct is a consequence of the fact that under $\left( \ref{prop-mod}%
\right) ,$ $\pi\left( X;\alpha^{\dagger}\right) =E\left( T|X\right) $ and
therefore 
\begin{align*}
& E\left[ V\left\{ T-\pi\left( X;\alpha^{\dagger}\right) \right\} \left\{
H\left( \beta^{\dagger}\right) -\Pi\left[ E\left\{ H\left(
\beta^{\dagger}\right) |X\right\} |V^{\dagger}\right] \right\} \right] \\
& =E\left[ V\left\{ T-E\left( T|X\right) \right\} H\left(
\beta^{\dagger}\right) \right] \\
& =E\left[ V\left\{ T-E\left( T|X\right) \right\} \left[ T\left\{
\gamma\left( X\right) -V^{T}\beta^{\dagger}\right\} +m_{0}\left( X\right) %
\right] \right] \\
& =E\left[ V\left\{ T-E\left( T|X\right) \right\} T\left( \gamma\left(
X\right) -V^{T}\beta^{\dagger}\right) \right]
\end{align*}
where the first and third equalities follow because $T-E\left( T|X\right) $
is uncorrelated with any function of $X$ and the second equality follows
after taking conditional expectations given $\left( T,X\right) $ inside the
expectation.

That $\left( \ref{main2}\right) $ holds when $\left( \ref{reg-mod}\right) $
is correct is a consequence of the fact that under $\left( \ref{reg-mod}%
\right) ,$\newline
$\Pi\left[ E\left\{ H\left( \beta^{\dagger}\right) |X\right\} |V^{\dagger}%
\right] =E\left\{ H\left( \beta^{\dagger}\right) |X\right\} $ and therefore%
\begin{align*}
& E\left[ V\left\{ T-\pi\left( X;\alpha^{\dagger}\right) \right\} \left\{
H\left( \beta^{\dagger}\right) -\Pi\left[ E\left\{ H\left(
\beta^{\dagger}\right) |X\right\} |V^{\dagger}\right] \right\} \right] \\
& =E\left[ V\left\{ T-\pi\left( X;\alpha^{\dagger}\right) \right\} \left\{
H\left( \beta^{\dagger}\right) -E\left\{ H\left( \beta^{\dagger }\right)
|X\right\} \right\} \right] \\
& =E\left[ V\left\{ T-E\left( T|X\right) \right\} \left\{ H\left(
\beta^{\dagger}\right) -E\left\{ H\left( \beta^{\dagger}\right) |X\right\}
\right\} \right] \\
& =E\left[ V\left\{ T-E\left( T|X\right) \right\} \left[ T\left\{
\gamma\left( X\right) -V^{T}\beta^{\dagger}\right\} +m_{0}\left( X\right)
-E\left\{ H\left( \beta^{\dagger}\right) |X\right\} \right] \right] \\
& =E\left[ V\left\{ T-E\left( T|X\right) \right\} T\left( \gamma\left(
X\right) -V^{T}\beta^{\dagger}\right) \right]
\end{align*}
where the second equality follows because $\left\{ H\left( \beta^{\dagger
}\right) -E\left\{ H\left( \beta^{\dagger}\right) |X\right\} \right\} $ is
uncorrelated with any function of $X,$ the third by conditioning on $\left(
T,X\right) $ insider the expectation and the fourth because $\left\{
T-E\left( T|X\right) \right\} $ is uncorrelated with any function of $X.$

\bigskip

\textbf{A.5. Proof of the identity} $\left( \ref{ts}\right) .$ By
definition, $\beta^{\dagger}V=\Pi_{\ \omega_{op}}\left[ \gamma\left(
X\right) |\left[ V\right] \right] .$ Consequently, 
\begin{align*}
\beta^{\dagger T}E\left( V\right) -{\tau}_{pop} & =-E\left\{ \Pi_{\
\omega_{op}}\left[ \gamma\left( X\right) |\left[ V\right] ^{\perp }\right]
\right\} \\
& =-E_{\omega_{op}}\left[ \left\{ Q+1\right\} \Pi_{\ \omega_{op}}\left[
\gamma\left( X\right) |\left[ V\right] ^{\perp}\right] \right] \\
& =E_{\omega_{op}}\left\{ \Pi_{\ \omega_{op}}\left[ Q+1|\left[ V\right]
^{\perp}\right] \Pi_{\ \omega_{op}}\left[ \gamma\left( X\right) |\left[ V%
\right] ^{\perp}\right] \right\} \\
& =E_{\omega_{op}}\left\{ \Pi_{\ \omega_{op}}\left[ Q|\left[ V\right]
^{\perp}\right] \Pi_{\ \omega_{op}}\left[ \gamma\left( X\right) |\left[ V%
\right] ^{\perp}\right] \right\}
\end{align*}
where the second equality is true by definition of $E_{\omega_{op}}\left(
\cdot\right) $ and $Q$, the third is true because $\Pi_{\ \omega_{op}}\left[
Q+1|\left[ V\right] \right] $ is orthogonal to $\Pi_{\ \omega_{op}}\left[
\gamma\left( X\right) |\left[ V\right] ^{\perp}\right] $ with respect to the
inner product $\,\left\langle ,\right\rangle _{\omega_{op}}$ and the fourth
is true because $1$ is the first component of the vector $V.$

\bigskip

\textbf{A.4. Proof that the identity }$\left( \ref{ts}\right) $\textbf{\
implies that }$\mathbb{P}_{n}\left( \widehat{\beta}^{T}V\right) $\textbf{\
is a DR estimator of }$\tau_{con}$\textbf{\textbf{\ and of} }$\tau_{pop}$%
\textbf{\ in model }$\mathcal{U}_{0}\left( \mathcal{F}\times\mathcal{P}%
\right) $\textbf{\ when }$\left( \ref{wop-mod}\right) $\textbf{\ holds.}

Suppose that $\left( \ref{wop-mod}\right) $ and model $\left( \ref{prop-mod}%
\right) $ hold and that $\widehat{\alpha}$ in formula $\left( \ref{betahat}%
\right) $\ is the ML estimator of $\alpha$ under this model. Then, $\mathbb{P%
}_{n}\left( \widehat{\beta}^{T}V\right) -{\tau}_{pop}\overset{P}{\rightarrow}%
0$ $\left( \text{and }\mathbb{P}_{n}\left( \widehat{\beta}^{T}V\right) -{\tau%
}_{con}\overset{P}{\rightarrow}0\right) $ because $\Pi_{\ \omega_{op}}\left[
Q|\left[ V\right] ^{\perp}\right] =0.$ On the other hand, if $\left( \ref%
{prop-mod}\right) $ does not hold but model $\left( \ref{reg-mod}\right) $
holds, then $\mathbb{P}_{n}\left( \widehat{\beta}^{T}V\right) -{\tau}_{pop}%
\overset{P}{\rightarrow}0$ $\left( \text{and }\mathbb{P}_{n}\left( \widehat{%
\beta}^{T}V\right) -{\tau}_{con}\overset{P}{\rightarrow}0\right) $ if, in
addition, model $\left( \ref{gamma}\right) $ holds, because then $\Pi_{\
\omega_{op}}\left[ \gamma\left( X\right) |\left[ V\right] ^{\perp}\right]
=0. $ We therefore conclude that $\mathbb{P}_{n}\left( \widehat{\beta}%
^{T}V\right) -{\tau }_{pop}\overset{P}{\rightarrow}0$ $\left( \text{and }%
\mathbb{P}_{n}\left( \widehat{\beta}^{T}V\right) -{\tau}_{con}\overset{P}{%
\rightarrow}0\right) $ when both models $\left( \ref{reg-mod}\right) $ and $%
\left( \ref{gamma}\right) $, or equivalently, when model $\left( \ref{total}%
\right) $ holds.

\bigskip

\textbf{A.6. Proof that} $\beta^{\dagger T}E_{\omega_{op}}\left( V\right)
=\tau_{pop,\omega_{op}}$

We have%
\begin{align*}
\beta^{\dagger T}E_{\omega_{op}}\left( V\right) & =E_{\omega_{op}}\left(
\beta^{\dagger T}V\right) \\
& =E_{\omega_{op}}\left\{ \Pi_{\omega_{op}}\left[ \gamma\left( X\right) |%
\left[ V\right] \right] \times1\right\} \\
& =E_{\omega_{op}}\left\{ \gamma\left( X\right) \times\Pi_{\omega_{op}} 
\left[ 1|\left[ V\right] \right] \right\} \\
& =E_{\omega_{op}}\left\{ \gamma\left( X\right) \right\} =\tau
_{pop,\omega_{op}}
\end{align*}
where the second equality is by $\beta^{\dagger T}V=\Pi_{\omega_{op}}\left[
\gamma\left( X\right) |\left[ V\right] \right] $, the third is because the
projections are computed with respect to the inner product, $\left\langle
A,B\right\rangle _{\omega_{op}}=E_{\omega_{op}}\left( AB\right) ,$ and the
fourth is because $V$ has first component equal to 1.

\textbf{A.7. Asymptotic distribution of the estimators in Section 4.}

Define the constants$,a=-E\left[ VV^{T}\pi\left( X\right) \left\{
1-\pi\left( X;\alpha^{\dagger}\right) \right\} \right] ,b=-E\left[
VV^{\dagger T}\left\{ \pi\left( X\right) -\pi\left( X;\alpha^{\dagger
}\right) \right\} \right] ,$

$c=-E\left( TV^{\dagger}V^{T}\right) ,d=-E\left( V^{\dagger}V^{\dagger
T}\right) ,e=-E\left\{ \left( Y-T\beta^{\dagger
T}V^{\dagger}-\theta^{\dagger T}V^{\dagger}\right) V\left\{ \left.
\partial\pi\left( X;\alpha\right) /\partial\alpha\right\vert
_{\alpha^{\dagger}}\right\} ^{T}\right\} ,$\newline
$k=E\left\{ \left. \partial S_{3}\left( \alpha\right)
/\partial\alpha\right\vert _{\alpha^{\dagger}}\right\} ,$ $\Sigma=\left\{
E_{\omega_{op}}\left( VV^{T}\right) \right\} ^{-1}$ and $\Omega=\left[
E\left\{ \omega_{op}\left( X\right) \right\} \right] ^{-1}.$ Note that when
model $\left( \ref{prop-mod}\right) $ holds, $b=0$ and when model $\left( %
\ref{reg-mod}\right) $ holds, $e=0.$ Define also, the random vectors%
\begin{equation*}
\psi\left( O\right) =\left( a-bd^{-1}c\right)
^{-1}S_{1}^{\dagger}-a^{-1}b\left( d-ca^{-1}b\right)
^{-1}S_{2}^{\dagger}-\left( a-bd^{-1}c\right) ^{-1}ek^{-1}S_{3}^{\dagger} 
\end{equation*}%
\begin{equation*}
\phi\left( O\right) =E_{\omega_{op}}\left( V^{T}\right) \psi\left( O\right)
+\Omega\omega_{op}\left( X\right) \left\{ V^{T}\beta^{\dagger
}-E_{\omega_{op}}\left( V^{T}\beta^{\dagger}\right) \right\} 
\end{equation*}%
\begin{equation*}
\varphi\left( O\right) =E\left( V^{T}\right) \psi\left( O\right) +\left\{
V^{T}\beta^{\dagger}-E\left( V^{T}\beta^{\dagger}\right) \right\} 
\end{equation*}
where $S_{j}^{\dagger}=S_{j}\left(
\beta^{\dagger},\alpha^{\dagger},\theta^{\dagger}\right) ,j=1,2,3$.\ Also,
let $\Sigma,\Omega,\Gamma _{\beta^{\dagger}},\Lambda_{\omega_{op},j}$ and $%
\Lambda_{j},j=1,2,3,$ be defined as in Section 4.4. The following Lemma
gives the asymptotic distribution of our estimators and concludes the proof of Theorem
1.

\textbf{Lemma A.1:} Under model $\mathcal{U}_{1}\left( \mathcal{F}\times%
\mathcal{P}\right) ,$

a) $\sqrt{n}\left( \widehat{\beta}-\beta^{\dagger}\right) \rightarrow
N\left( 0,\Psi\right) $ where $\Psi=var\left\{ \psi\left( O\right) \right\}
. $ When both models $\left( \ref{prop-mod}\right) $ and $\left( \ref%
{reg-mod}\right) $ hold and $var\left( Y|X,T\right) =\sigma^{2}$, we have
that $\psi\left( O\right) =\Omega\Sigma S_{1}^{\dagger}$ and $%
\Psi=\Omega\Sigma\Gamma_{\beta^{\dagger}}\Sigma$ where $\Gamma_{\beta
^{\dagger}}$ was defined in Section 4.4.

b) $\sqrt{n}\left\{ \mathbb{P}_{n,\omega_{op}}\left( \widehat{\beta}%
^{T}V\right) -E_{\omega_{op}}\left( \beta^{\dagger T}V\right) \right\}
\rightarrow N\left( 0,\Lambda_{\omega_{op}}\right) $ where $\Lambda
_{\omega_{op}}=var\left\{ \phi\left( O\right) \right\} .$ When both models $%
\left( \ref{prop-mod}\right) $ and $\left( \ref{reg-mod}\right) $ hold and $%
var\left( Y|X,T\right) =\sigma^{2}$, we have that $\Lambda_{\omega
_{op}}=\Lambda_{\omega_{op},1}+\Lambda_{\omega_{op},2}+\Lambda_{%
\omega_{op},3}$

c) $\sqrt{n}\left\{ \mathbb{P}_{n}\left( \widehat{\beta}^{T}V\right)
-E\left( \beta^{\dagger T}V\right) \right\} \rightarrow N\left(
0,\Lambda\right) $ where $\Lambda=var\left\{ \varphi\left( O\right) \right\}
.$ When both models $\left( \ref{prop-mod}\right) $ and $\left( \ref{reg-mod}%
\right) $ hold, $\Lambda=\Lambda_{1}+\Lambda_{2}+\Lambda_{3}.$

d) If both models $\left( \ref{prop-mod}\right) $ and $\left( \ref{reg-mod}%
\right) $ hold, $var\left( Y|X,T\right) =\sigma^{2}$ and in addition, model $%
\left( \ref{gamma}\right) $ for $\gamma\left( X\right) $ holds, we have that 
$\Lambda_{\omega_{op},3}=\Lambda_{3}=0$ and $\Lambda _{\omega_{op}}$ and $%
\Lambda$ are as defined in Theorem 1 respectively.

\textbf{Proof: }Let $\widehat{\rho}\equiv\left( \widehat{\beta}^{T},\widehat{%
\theta}^{T},\widehat{\alpha}^{T}\right) ^{T},\rho^{\dag}\equiv\left(
\beta^{\dag},\theta^{\dag},\alpha^{\dag}\right) .$ We have already shown in
A.2 that under $\mathcal{U}_{1}\left( \mathcal{F}\times\mathcal{P}\right) ,%
\widehat{\rho}\overset{P}{\rightarrow}\rho^{\dag}.$ A Taylor's expansion
gives 
\begin{equation*}
\sqrt{n}\mathbb{P}_{n}\left\{ S\left( \widehat{\rho}\right) \right\} =0=%
\sqrt{n}\mathbb{P}_{n}\left\{ S\left( \rho^{\dag}\right) \right\} +\frac{%
\partial}{\partial\rho^{T}}\left. \mathbb{P}_{n}\left\{ S\left( \rho\right)
\right\} \right\vert _{\overline{\rho}}\sqrt{n}\left( \widehat{\rho}%
-\rho^{\dag}\right) 
\end{equation*}
for some $\overline{\rho}=\left( \overline{\beta}^{T},\overline{\theta}^{T},%
\overline{\alpha}^{T}\right) ^{T}$ such that $\left\Vert \rho^{\dag }-%
\overline{\rho}\right\Vert \leq\left\Vert \rho^{\dag}-\widehat{\rho }%
\right\Vert .$ A uniform Law of Large Numbers now gives 
\begin{equation}
\sqrt{n}\left( \widehat{\rho}-\rho^{\dag}\right) =-M^{-1}\sqrt{n}\mathbb{P}%
_{n}\left\{ S\left( \rho^{\dag}\right) \right\} +o_{p}\left( 1\right)
\label{kk}
\end{equation}
where 
\begin{equation*}
M=\left( 
\begin{array}{ccc}
a & b & e \\ 
c & d & 0 \\ 
0 & 0 & k%
\end{array}
\right) 
\end{equation*}
is equal to $E\left\{ \left. \frac{\partial}{\partial\rho^{T}}S\left(
\rho\right) \right\vert _{\rho^{\dag}}\right\} .$ Using formulae for the
inversion of a partitioned matrix it can be verified that 
\begin{equation*}
M^{-1}=\left( 
\begin{array}{ccc}
(a-bd^{-1}c)^{-1} & -a^{-1}b(d-ca^{-1}b)^{-1} & -(a-bd^{-1}c)^{-1}ek^{-1} \\ 
-d^{-1}c(a-bd^{-1}c)^{-1} & (d-ca^{-1}b)^{-1} & 
d^{-1}c(a-bd^{-1}c)^{-1}ek^{-1} \\ 
0 & 0 & k^{-1}%
\end{array}
\right) 
\end{equation*}
Replacing this formula in $\left( \ref{kk}\right) $ we obtain, in
particular, that $\sqrt{n}\left( \widehat{\beta}-\beta^{\dag}\right) =-\sqrt{%
n}\mathbb{P}_{n}\left\{ \psi\left( O\right) \right\} +o_{p}\left( 1\right) .$
The Central Limit Theorem now implies that $\sqrt{n}\left( \widehat{\beta}%
-\beta^{\dag}\right) \rightarrow N\left( 0,\Psi\right) $. When models $%
\left( \ref{prop-mod}\right) $ and $\left( \ref{reg-mod}\right) $ hold, $%
b=0, $ $e=0$ and $a^{-1}=\Omega\Sigma.$ Consequently, $\psi\left( O\right)
=\Omega\Sigma S_{1}^{\dagger}$ and $var\left\{ \psi\left( O\right) \right\}
=\Omega^{2}\Sigma var\left( S_{1}^{\dagger }\right) \Sigma.$ If, in
addition, var$\left( Y|T,X\right) =\sigma^{2}$ we obtain 
\begin{align*}
& \Omega^{2}var\left( S_{1}^{\dagger}\right) \\
& =\Omega^{2}E\left\{ var\left( S_{1}^{\dagger}|T,X\right) \right\}
+\Omega^{2}var\left\{ E\left( S_{1}^{\dagger}|T,X\right) \right\} \\
& =\Omega^{2}\sigma^{2}E\left\{ \left[ VV^{T}\left\{ T-\pi\left( X\right)
\right\} ^{2}\right] \right\} \\
& +\Omega^{2}var\left[ V\left\{ T-\pi\left( X\right) \right\} ^{2}\left\{
\gamma\left( X\right) -\Pi_{\omega_{op}}\left[ \gamma\left( X\right) |\left[
V\right] \right] \right\} \right] \\
& =\sigma^{2}\Omega E_{\omega_{op}}\left( VV^{T}\right) +\Omega^{2}E\left[
VV^{T}\left\{ T-\pi\left( X\right) \right\} ^{4}\left\{ \Pi_{\omega_{op}}%
\left[ \gamma\left( X\right) |\left[ V\right] ^{\perp}\right] \right\} ^{2}%
\right] \\
& =\sigma^{2}\Omega E_{\omega_{op}}\left( VV^{T}\right) +\Omega
E_{\omega_{op}}\left[ VV^{T}b\left( X\right) \left\{ \Pi_{\omega_{op}}\left[
\gamma\left( X\right) |\left[ V\right] ^{\perp}\right] \right\} ^{2}\right]
\end{align*}
thus showing part a) of Theorem 1.

Part ii) follows from%
\begin{align*}
& \sqrt{n}\left\{ \mathbb{P}_{n,\omega_{op}}\left( \widehat{\beta}%
^{T}V\right) -E_{\omega_{op}}\left( \beta^{\dagger T}V\right) \right\} \\
& =\sqrt{n}\left( \widehat{\beta}-\beta^{\dagger}\right) ^{T}\mathbb{P}%
_{n,\omega_{op}}\left( V\right) \\
& +\sqrt{n}\mathbb{P}_{n}\left[ \omega_{op}\left( X\right) \left\{
\beta^{\dagger T}V-E_{\omega_{op}}\left( \beta^{\dagger T}V\right) \right\} %
\right] \left[ \mathbb{P}_{n}\left\{ \omega_{op}\left( X\right) \right\} %
\right] ^{-1} \\
& =\sqrt{n}E_{\omega_{op}}\left( V^{T}\right) \mathbb{P}_{n}\left\{
\psi\left( O\right) \right\} +\Omega\sqrt{n}\mathbb{P}_{n}\left[
\omega_{op}\left( X\right) \left\{ \beta^{\dagger T}V-E_{\omega_{op}}\left(
\beta^{\dagger T}V\right) \right\} \right] +o_{p}\left( 1\right) \\
& =\sqrt{n}\mathbb{P}_{n}\left\{ \phi\left( O\right) \right\} +o_{p}\left(
1\right)
\end{align*}
When models $\left( \ref{prop-mod}\right) $ and $\left( \ref{reg-mod}\right) 
$ hold we have

\begin{align*}
\phi\left( O\right) & =\Omega E_{\omega_{op}}\left( V^{T}\right) \Sigma
V\left\{ H\left( \beta^{\dagger}\right) -\theta^{\dagger T}V^{\dagger
}\right\} \left\{ T-\pi\left( X\right) \right\} \\
& +\Omega\omega_{op}\left( X\right) \left\{
V^{T}\beta^{\dagger}-E_{\omega_{op}}\left( V^{T}\beta^{\dagger}\right)
\right\} \\
& =\Omega\left[ \left\{ H\left( \beta^{\dagger}\right) -\theta^{\dagger
T}V^{\dagger}\right\} \left\{ T-\pi\left( X\right) \right\} +\omega
_{op}\left( X\right) \left\{ V^{T}\beta^{\dagger}-E_{\omega_{op}}\left(
V^{T}\beta^{\dagger}\right) \right\} \right]
\end{align*}
where the third equality follows because $E_{\omega_{op}}\left( V^{T}\right)
\Sigma V=E_{\omega_{op}}\left( 1V^{T}\right) E_{\omega_{op}}\left(
VV^{T}\right) V=\Pi_{\omega_{op}}\left[ 1|\left[ V\right] \right] =1$ since $%
1$ is the first component of the vector $V.$ If in addition, $var\left(
Y|T,X\right) =\sigma^{2}$ we have%
\begin{align*}
& var\left\{ \phi\left( O\right) \right\} \\
& =E\left[ var\left\{ \phi\left( O\right) |T,X\right\} \right] +var\left[
E\left\{ \phi\left( O\right) |T,X\right\} \right] \\
& =\Omega^{2}E\left[ \sigma^{2}\left\{ T-\pi\left( X\right) \right\} ^{2}%
\right] \\
& +\Omega^{2}var\left[ \left\{ T-\pi\left( X\right) \right\} ^{2}\left\{
\gamma\left( X\right) -\Pi_{\omega_{op}}\left[ \gamma\left( X\right) |\left[
V\right] \right] \right\} \right. \\
& \left. +\omega_{op}\left( X\right) \left\{ V^{T}\beta^{\dagger
}-E_{\omega_{op}}\left( V^{T}\beta^{\dagger}\right) \right\} \right] \\
& =\Omega\sigma^{2}+\Omega^{2}E\left[ \left\{ T-\pi\left( X\right) \right\}
^{4}\left\{ \Pi_{\omega_{op}}\left[ \gamma\left( X\right) |\left[ V\right]
^{\perp}\right] \right\} ^{2}\right] \\
& +\Omega^{2}E\left[ \omega_{op}\left( X\right) ^{2}\left\{
V^{T}\beta^{\dagger}-E_{\omega_{op}}\left( V^{T}\beta^{\dagger}\right)
\right\} ^{2}\right] \\
& +\Omega^{2}E\left[ \left\{ T-\pi\left( X\right) \right\}
^{2}\Pi_{\omega_{op}}\left[ \gamma\left( X\right) |\left[ V\right] ^{\perp }%
\right] \omega_{op}\left( X\right) \left\{
V^{T}\beta^{\dagger}-E_{\omega_{op}}\left( V^{T}\beta^{\dagger}\right)
\right\} \right] \\
& =\Omega\sigma^{2}+\Omega E_{\omega_{op}}\left[ b\left( X\right) \left\{
\Pi_{\omega_{op}}\left[ \gamma\left( X\right) |\left[ V\right] ^{\perp }%
\right] \right\} ^{2}\right] \\
& +\Omega E_{\omega_{op}}\left[ \left\{ V^{T}\beta^{\dagger}-E_{\omega
_{op}}\left( V^{T}\beta^{\dagger}\right) \right\} ^{2}\right] \\
& +2\Omega E_{\omega_{op}}\left[ \omega_{op}\left( X\right) \Pi
_{\omega_{op}}\left[ \gamma\left( X\right) |\left[ V\right] ^{\perp }\right]
\left\{ V^{T}\beta^{\dagger}-E_{\omega_{op}}\left(
V^{T}\beta^{\dagger}\right) \right\} \right]
\end{align*}
The last member is precisely $\Lambda_{\omega_{op}},$ thus finalizing the
proof of part i) of the Lemma and part b) of Theorem 1.

To show part iii) write

\begin{align*}
& \sqrt{n}\left\{ \mathbb{P}_{n}\left( \widehat{\beta}^{T}V\right) -E\left(
\beta^{\dagger T}V\right) \right\} \\
& =\sqrt{n}\left( \widehat{\beta}-\beta^{\dagger}\right) ^{T}\mathbb{P}%
_{n}\left( V\right) +\sqrt{n}\mathbb{P}_{n}\left[ \left\{ \beta^{\dagger
T}V-E\left( \beta^{\dagger T}V\right) \right\} \right] \\
& =\sqrt{n}E\left( V^{T}\right) \mathbb{P}_{n}\left\{ \psi\left( O\right)
\right\} +\sqrt{n}\mathbb{P}_{n}\left\{ \beta^{\dagger T}V-E\left(
\beta^{\dagger T}V\right) \right\} +o_{p}\left( 1\right) \\
& =\sqrt{n}\mathbb{P}_{n}\left\{ \varphi\left( O\right) \right\}
+o_{p}\left( 1\right)
\end{align*}
When models $\left( \ref{prop-mod}\right) $ and $\left( \ref{reg-mod}\right) 
$ hold and $var\left( Y|T,X\right) =\sigma^{2}$ we have%
\begin{align*}
& var\left\{ \varphi\left( O\right) \right\} \\
& =E\left[ var\left\{ \varphi\left( O\right) |T,X\right\} \right] +var\left[
E\left\{ \varphi\left( O\right) |T,X\right\} \right] \\
& =\sigma^{2}\Omega^{2}E\left[ E\left( V^{T}\right) \Sigma VV^{T}\Sigma
E\left( V\right) \left\{ T-\pi\left( X\right) \right\} ^{2}\right] \\
& +\Omega^{2}var\left[ E\left( V^{T}\right) \Sigma V\left\{ T-\pi\left(
X\right) \right\} ^{2}\Pi_{\omega_{op}}\left[ \gamma\left( X\right) |\left[ V%
\right] ^{\bot}\right] +\left\{ V^{T}\beta^{\dagger}-E\left(
V^{T}\beta^{\dagger}\right) \right\} \right] \\
& =\sigma^{2}E\left[ \left\{ \Pi_{\omega_{op}}\left[ \omega_{op}\left(
X\right) ^{-1}|\left[ V\right] \right] \right\} ^{2}\omega_{op}\left(
X\right) \right] \\
& +var\left[ \Pi_{\omega_{op}}\left[ \omega_{op}\left( X\right) ^{-1}|\left[
V\right] \right] \omega_{op}\left( X\right) \Pi_{\omega _{op}}\left[
\gamma\left( X\right) |\left[ V\right] ^{\bot}\right] +\left\{
V^{T}\beta^{\dagger}-E\left( V^{T}\beta^{\dagger}\right) \right\} \right] \\
& =\sigma^{2}\Omega^{-1}E_{\omega_{op}}\left[ \left\{ \Pi_{\omega_{op}}\left[
\omega_{op}\left( X\right) ^{-1}|\left[ V\right] \right] \right\} ^{2}\right]
\\
& +\Omega^{-1}E_{\omega_{op}}\left[ \left\{ \Pi_{\omega_{op}}\left[
\omega_{op}\left( X\right) ^{-1}|\left[ V\right] \right] \right\}
^{2}\left\{ \Pi_{\omega_{op}}\left[ \gamma\left( X\right) |\left[ V\right]
^{\bot}\right] \right\} ^{2}\right] \\
& +E\left[ \left\{ V^{T}\beta^{\dagger}-E_{\omega_{op}}\left(
V^{T}\beta^{\dagger}\right) \right\} ^{2}\right] \\
& +2\Omega^{-1}E_{\omega_{op}}\left\{ \Pi_{\omega_{op}}\left[ \omega
_{op}\left( X\right) ^{-1}|\left[ V\right] \right] \Pi_{\omega_{op}}\left[
\gamma\left( X\right) |\left[ V\right] ^{\bot}\right] \left\{
V^{T}\beta^{\dagger}-E_{\omega_{op}}\left( V^{T}\beta^{\dagger}\right)
\right\} \right\}
\end{align*}
where the third equality follows because $\Omega E\left( V^{T}\right) \Sigma
V=\Pi_{\omega_{op}}\left[ \omega_{op}\left( X\right) ^{-1}|\left[ V\right] %
\right] .$ This concludes the proof of part iii) of the Lemma and part c) of
Theorem 1.

Part iv) of the Lemma and d) of the Theorem 1 is immediate upon noticing
that when model $\left( \ref{gamma}\right) $ holds, $\Pi_{\omega_{op}}\left[
\gamma\left( X\right) |\left[ V\right] ^{\bot}\right] =0$ and $%
V^{T}\beta^{\dagger}=\gamma\left( X\right) .$

\end{document}